\documentclass[11pt]{article}

\usepackage{amsmath,amsthm,amssymb,fullpage,graphics,hyperref,mathrsfs,epsfig,graphicx}
\usepackage[abbrev]{amsrefs}
\usepackage{txfonts}
  \ifx \LabelFigloaded\MYundefined
  \else
    \immediate\write16{ !!! LabelFig.tex ALREADY loaded.}
   \fi
 
  \def\LabelFigloaded{\relax}% now loaded

  \chardef\LabelFigCatAt\the\catcode`\@
  \catcode`\@=11
 
 \let\LabelFigwlog@ld\wlog
 \def\wlog#1{\relax}

 \ifx\\\MYundefined@
    \let\\\relax
 \fi
 
 \def\N@wif{\csname newif\endcsname }
 \def\Temp@ {\N@wif\ifIN@}
 \ifx\INN@\MYundefined@
    \else \let\Temp@\relax
 \fi
 \Temp@
 
  \def\IN@{\expandafter\INN@\expandafter}
  \long\def\INN@0#1@#2@{\long\def\NI@##1#1##2##3\ENDNI@
    {\ifx\m@rker##2\IN@false\else\IN@true\fi}%
     \expandafter\NI@#2@@#1\m@rker\ENDNI@}
  \def\m@rker{\m@@rker}
 
 \def\Shifted@@#1#2#3{\setbox0=\hbox{#3}%
   \raise -\dp0\vbox {\kern-#2%
       \hbox {\kern#1\box0\kern-#1}%
           \kern#2}}

 \newcount\gridcount
 \newbox\auxGridbox@ \newbox\hGridbox@ \newbox\vGridbox@
 \newbox\Labelbox@ \newbox\auxLabelbox@
 \newbox\Coordinatebox@
 \newtoks\Labeltoks@
 \newdimen\Wdd@ \newdimen\Htt@
 
 \def\hRule@{\advance\gridcount -2%
   \vskip-.2pt\hrule\vskip-.2pt\vfil
   \llap{\smash{\raise -2.5pt
     \hbox{.\number\gridcount\kern2pt}}}%
           \vskip-.2pt\hrule\vskip-.2pt\vfil}
 
 \def\vRule@{\advance\gridcount 2%
   \hskip-.2pt\vrule\hskip-.2pt\hfil
   \setbox\auxGridbox@=\vbox to 0pt
      {\vskip \Htt@\vskip 2pt
        \hbox{\kern-3.5pt.\number\gridcount}\vss}%
      \wd\auxGridbox@=0pt \box\auxGridbox@
   \hskip-.2pt\vrule\hskip-.2pt\hfil}
 
 \def\PlaceGrid@@{\gridcount=10%
  \setbox\hGridbox@=%
    \hbox{\hbox{\GridSpider@{\hskip-.4pt\vrule}%
             \vbox to \Htt@{\offinterlineskip\parindent=\z@%
                \GridSpider@{\vskip-.4pt\hrule}\vfil
                \hRule@\hRule@\hRule@\hRule@
                  \vskip-.2pt\hrule\vskip-.2pt\vfil
                \hbox to \Wdd@{\hfil}%
             \GridSpider@{\hrule\vskip-.4pt}}%
         \GridSpider@{\vrule\hskip-.4pt}}}%
  \gridcount=0%
  \setbox\vGridbox@=
     \hbox{\vbox{\offinterlineskip\parindent=0pt\hsize=0pt
       \GridSpider@{\vskip-.4pt\hrule}%
             \hbox to \Wdd@{%
                \GridSpider@{\hskip-.4pt\vrule}\hfil
                \vtop to \Htt@{\vfil}%
                     \vRule@\vRule@\vRule@\vRule@
                     \hskip-.2pt\vrule\hskip-.2pt\hfil
             \GridSpider@{\vrule\hskip-.4pt}}%
         \GridSpider@{\hrule\vskip-.4pt}}}%
  \wd\hGridbox@=0pt\ht\hGridbox@=0pt
  \wd\vGridbox@=0pt\ht\vGridbox@=0pt
 \hbox{\box\hGridbox@\box\vGridbox@}%
  }
 
 \def\SetLabels#1\endSetLabels{%
   \Labeltoks@={#1}}
 
 \def\GridSpider@#1{#1}
 \let\PlaceGrid@\relax
 \def\ShowGrid{\let\PlaceGrid@\PlaceGrid@@}
 
 \def\bAdjust@@{%
     \setbox\auxLabelbox@=\hbox{\raise \dp\auxLabelbox@
            \box\auxLabelbox@}}
 \def\bAdjust@{\let\vAdjust@\bAdjust@@}
 
 \def\tAdjust@@{%
     \setbox\auxLabelbox@=\hbox{\raise-\ht\auxLabelbox@
            \box\auxLabelbox@}}
 \def\tAdjust@{\let\vAdjust@\tAdjust@@}
 
 \let\vAdjust@\relax
 
 \def\lAdjust@{\let\hAdjust@\rlap}
 \def\rAdjust@{\let\hAdjust@\llap}
 
 \let\hAdjust@\relax\let\vAdjust@\relax
 
 \def\FetchLabel@#1(#2*#3)#4\\#5\endFetchLabel@{%
     \ignorespaces#1\unskip
     \Labeltoks@={#5}%
     \setbox\auxLabelbox@=%
        \hbox to 0pt{\hss\ignorespaces\hAdjust@
          {\ignorespaces#4\unskip}\hss}%
     \vAdjust@
     \let\hAdjust@\relax\let\vAdjust@\relax
     \setbox\Labelbox@=\hbox to 0pt{%
       \box\Labelbox@
       \Shifted@@{#2\Wdd@}{#3\Htt@}{\box\auxLabelbox@}}%
     \ht\Labelbox@=0pt\dp\Labelbox@=0pt
     }
 
 \def\PlaceLabels@@{\bgroup\def\Cr@{\\}%
     \let\L\lAdjust@\let\R\rAdjust@
     \let\B\bAdjust@\let\T\tAdjust@
     \loop
     \IN@0\Cr@ @\the\Labeltoks@ @\relax
     \ifIN@ \expandafter
       \FetchLabel@\the\Labeltoks@\endFetchLabel@
     \repeat
     \box\Labelbox@\egroup}%
 
 \let \PlaceLabels@\PlaceLabels@@
 
 \def\AffixLabels#1{\setbox\Coordinatebox@=\hbox{#1}%
      \Wdd@=\wd\Coordinatebox@ \Htt@=\ht\Coordinatebox@
      \advance\Htt@ \dp \Coordinatebox@
      \hbox{\copy\Coordinatebox@\kern-\Wdd@%
           \Shifted@@{0pt}{-\dp\Coordinatebox@}%
            {\PlaceGrid@\PlaceLabels@}%
           \kern\Wdd@}}
 
   \let\wlog\LabelFigwlog@ld %%restore logging
   \catcode`\@=\LabelFigCatAt  %%12 or 13

\newcommand{\C}{\mathscr{C}}

\newcommand{\E}{\mathbb{E}}
\newcommand{\R}{\mathbb{R}}
\newcommand{\D}{\mathscr{D}}
\newcommand{\W}{\mathscr{W}}

\DeclareMathOperator{\diam}{diam}
\DeclareMathOperator{\lca}{lca}

\DeclareMathOperator{\cost}{cost}

\newcommand{\MET}{\mathrm{MET}}
\newcommand{\Opt}{\mathrm{Opt}}
\newcommand{\rd}{\mathrm{rd}}
\newtheorem{theorem}{Theorem}
\newtheorem{proposition}[theorem]{Proposition}
\newtheorem{lemma}[theorem]{Lemma}
\newtheorem{claim}[theorem]{Claim}

\renewcommand{\le}{\leqslant}
\renewcommand{\ge}{\geqslant}

\renewcommand{\geq}{\geqslant}

\theoremstyle{definition}

\theoremstyle{remark}
\newtheorem{question}{Open Question}
\newtheorem{remark}[question]{Remark}

\newcommand{\T}{\mathcal T}
\newcommand{\N}{\mathcal{N}}

\newcommand{\Z}{\mathbb Z}
\newcommand{\e}{\varepsilon}

\newcommand{\Lip}{\mathrm{Lip}}

\newcommand{\poly}{\mbox{poly}}

\begin{document}

\title{Maximum gradient embeddings and monotone clustering}
\author{Manor Mendel\\Computer Science Division\\
The Open University of Israel\\ {\tt mendelma@gmail.com} \and Assaf
Naor\\ Courant Institute\\ New York University\\{\tt
naor@cims.nyu.edu} }
\date{}
\maketitle

\begin{abstract} Let $(X,d_X)$ be an $n$-point metric space. We show that
there exists a distribution $\D$ over non-contractive embeddings
into trees $f:X\to T$  such that for every $x\in X$,
$$
 \E_\D \left[\max_{y\in X\setminus\{x\}}
\frac{d_T(f(x),f(y))}{d_X(x,y)}\right] \le C (\log n)^2,
$$
where $C$ is a universal constant. Conversely we show that the
above quadratic dependence on $\log n$ cannot be improved in
general. Such embeddings, which we call {\em maximum gradient
embeddings}, yield a
 framework for the design of approximation algorithms for a wide
 range of
clustering problems with monotone costs, including fault-tolerant
versions of $k$-median and facility location.
\end{abstract}

%\thispagestyle{empty}

%\newpage

%\setcounter{page}{1}

\section{Introduction}

Metric embeddings are an invaluable tool in analysis, Riemannian
geometry, group theory, graph theory, and the design of
approximation algorithms. In most cases embeddings are used to
``simplify" a geometric object that we wish to understand, or on
which we need to perform certain algorithmic tasks. Thus one tries
to faithfully represent a metric space as a subset of another space
with controlled geometry, whose structure is well enough understood
to successfully address the problem at hand. There is some obvious
flexibility in this approach: Both the choice of target space and
the notion of faithfulness of an embedding can be adapted to the
problem that we wish to solve. Of course, once these choices are
made, the main difficulty is the construction of the required
embedding, and in the algorithmic context we have the additional
requirement that the embedding can be computed efficiently.

%The present paper is inspired by  problems from mathematical
%analysis, but its main motivation is algorithmic.
In this paper we introduce a new notion of embedding, called {\em
maximum gradient embeddings}, which turns out to be perfectly suited
for approximating a wide range of clustering problems. We then
provide optimal maximum gradient embeddings of general finite metric
spaces, and use them to design
 approximation algorithms for several
clustering problems.  These embeddings yield a generic approach to
many problems, and we give some examples that illustrate this fact.

%The flexibility that was described above in the choice of notions
%of embedding  has been exploited to great success by numerous
%authors in the past four decades. Due to the vast amount of work
%on this topic, we will not give references to the many embedding
%notions that were used in the mathematical and computer science
%literature. We wish to stress, however, that apart from the most
%studied problem of bi-Lipschitz embeddings into Euclidean space,
%there are a lot of variants of this problem which are useful in
%many different contexts in mathematics and computer science. These
%notions include $L_1$ embeddings, embeddings into dominated $L_1$
%metrics, low dimensional embeddings and dimension reduction,
%volume preserving embeddings, Fr\'echet embeddings, snowflake,
%quasi-isometric, coarse, uniform, and quasisymmetric embeddings,
%embeddings into hyperbolic spaces, Tits buildings and symmetric
%spaces, embeddings of subsets, quotients, subsets of quotients and
%quotients of subsets, embeddings into distributions over trees,
%ultrametrics and spanning trees, multi-embeddings, embeddings with
%slack, nearest neighbor preserving embeddings, spanners, additive
%distortion, and various notions of average distortion. In this
%paper the target spaces will be distributions over tree metrics
%(also known as products of trees in the mathematical literature).
%Our notion of faithfulness is new, and will be described below.

Due to their special structure, it is natural to try to embed metric
spaces into trees. This is especially important for algorithmic
purposes, as many hard problems are tractable on trees.
Unfortunately, this is too much to hope for in the bi-Lipschitz
category: As shown by Rabinovich and Raz~\cite{RR98} the $n$-cycle
incurs distortion $\Omega(n)$ in any embedding into a tree. However,
one can relax this idea and look for a {\em random} embedding into a
tree which is faithful on average.

%Such an approach has been developed in recent years by
%mathematicians and computer scientists. In the mathematical
%literature this is referred to as embeddings into products of trees,
%and it is an invaluable tool in the study of negatively curved
%spaces (see for example~\cites{Dra03,BS05,NPSS05}).

Randomized embeddings into trees via mappings which do not contract
distances (also known as probabilistic embeddings into dominating
trees) became an important algorithmic paradigm due to the work of
Bartal~\cites{Bartal96,Bartal98} (see also~\cites{AKPW95,EEST05} for
the related problem of embedding graphs into distributions over
spanning trees). This work led to the design of many approximation
algorithms for a wide range of NP hard problems. In some cases the
best known approximation factors are due to the ``probabilistic
tree" approach, while in other cases improved algorithms have been
subsequently found after the original application of probabilistic
embeddings was discovered. But, in both cases it is clear that the
strength of Bartal's approach is that it is generic: For a certain
type of problem one can quickly get a polylogarithmic approximation
using probabilistic embedding into trees, and then proceed to
analyze certain particular cases if one desires to find better
approximation guarantees. However, probabilistic embeddings into
trees do not always work. In~\cite{BM04} Bartal and Mendel
introduced the weaker notion of multi-embeddings, and used it to
design improved algorithms for special classes of metric spaces.
Here we {\em strengthen} this notion to maximum gradient embeddings,
yielding a faithfulness measure which is nevertheless weaker than
bi-Lipschitz, and use it to design approximation algorithms for
harder problems to which regular probabilistic embeddings do not
apply.

%From a mathematical viewpoint, analyzing this stronger notion of
%embedding (which is nevertheless weaker than bi-Lipschitz embedding)
%creates new challenges, and arguably leads to some surprising
%results.

Let $(X,d_X)$ and $(Y,d_Y)$ be metric spaces, and fix a mapping
$f:X\to Y$. We shall say that $f$ is {\em non-contractive} if for
every $x,y\in X$ we have $d_Y(f(x),f(y))\ge d_X(x,y)$. The {\em
maximum gradient} of $f$ at a point $x\in X$ is defined as
\begin{eqnarray}\label{def:gradi}
|\nabla f(x)|_\infty=\sup_{y\in X\setminus\{x\}}
\frac{d_Y(f(x),f(y))}{d_X(x,y)}.
\end{eqnarray}
Thus the {\em Lipschitz constant} of $f$ is given by
$$
\|f\|_{\Lip}=\sup_{x\in X}|\nabla f(x)|_\infty.
$$
Note that in the mathematical literature, mostly in the context of
the study of isoperimetry on general geodesic metric measure
spaces (see for example~\cites{BH97,Led01}), it is common to define
the {\em modulus of the gradient} of $f$ at $x\in X$ as
\begin{eqnarray}\label{def:modulus}
|\nabla f(x)|=\limsup_{y\to x} \frac{d_Y(f(x),f(y))}{d_X(x,y)}.
\end{eqnarray}
The definition in~\eqref{def:modulus} is very natural in the
context of connected metric spaces, but in the context of finite
metric spaces it clearly makes more sense to deal with the maximum
gradient as defined in~\eqref{def:gradi}.

In what follows when we refer to a tree metric we mean the
shortest-path metric on a graph-theoretical tree with weighted
edges. Recall that $(U,d_U)$ is an ultrametric if for every
$u,v,w\in U$ we have $d_U(u,v)\le \max\{d_U(u,w),d_U(w,v)\}$. It
is well known that ultrametrics are tree metrics. The following
result is due to Fakcharoenphol, Rao and Talwar~\cite{FRT03}, and
is a slight improvement over an earlier theorem of
Bartal~\cite{Bartal98}. For every $n$-point metric space $(X,d_X)$
there is a distribution $\D$ over non-contractive embeddings into
ultrametrics $f:X\to U$ such that
\begin{eqnarray}\label{eq:bartal}
\max_{\substack{x,y\in X\\x\neq y}}\ \E_\D\left[
\frac{d_U(f(x),f(y)}{d_X(x,y)}\right]=O(\log n).
\end{eqnarray}
The logarithmic upper bound in~\eqref{eq:bartal} cannot be
improved in general.

Inequality~\eqref{eq:bartal} is extremely useful for optimization
problems whose objective function is linear in the distances,
since by linearity of expectation it reduces such tasks to trees,
with only a logarithmic loss in the approximation guarantee. When
it comes to non-linear problems, the use of~\eqref{eq:bartal} is
very limited. We will show that this issue can be addressed using
the following theorem, which is our main result.

\begin{theorem}\label{thm:ourmax} Let $(X,d_X)$ be an $n$-point
metric space. Then there exists a distribution $\D$ over
non-contractive embeddings into ultrametrics $f:X\to U$ (thus both
the ultrametric $(U,d_U)$ and the mapping $f$ are random) such that
for every $x\in X$,
$$
\E_\D \left[|\nabla f(x)|_\infty\right] \le C (\log n)^2,
$$
where $C$ is a universal constant.

 On the other hand there exists
a universal constant $c>0$ and arbitrarily large $n$-point metric
spaces $Y_n$ such that for any distribution over non-contractive
embeddings into trees $f:Y_n\to T$ there is necessarily some $x\in
Y_n$ for which
$$
\E_\D \left[|\nabla f(x)|_\infty\right] \ge c (\log n)^2.
$$
\end{theorem}

We call embeddings as in Theorem~\ref{thm:ourmax}, i.e. embeddings
with small expected maximum gradient, {\em maximum gradient
embeddings into distributions over trees} (in what follows we will
only deal with distributions over trees, so we will drop the last
part of this title when referring to the embedding, without creating
any ambiguity). The proof of the upper bound in
Theorem~\ref{thm:ourmax} is a modification of an argument of
Fakcharoenphol, Rao and Talwar~\cite{FRT03}, which is based on ideas
from~\cites{Bartal96,CKR04}. It uses the same stochastic
decomposition of metric spaces as in~\cite{FRT03}, but it relies on
properties of it which are well known to experts, yet  have not been
exploited in full strength in previous applications. The argument
appears in Section~\ref{sec:proof theorem}. Alternative proofs of
the main technical step of the proof of the upper bound in
Theorem~\ref{thm:ourmax} can be also deduced from the results
of~\cite{MN05} or an argument in the proof of Lemma 2.1
in~\cite{GHR06}. In both of these references the required inequality
is deduced from an improved analysis of the specific stochastic
decomposition of Calinescu, Karloff and Rabani~\cite{CKR04} that was
used in~\cite{FRT03}. Here we present a different approach, which
shows that the ``padding inequality" proved by Fakcharoenphol, Rao
and Talwar in~\cite{FRT03} can be used as a ``black box" to yield a
maximum gradient embedding, and there is no need to recall how the
stochastic decomposition was originally defined.

%Our proof can be viewed as a variant of Talagrand's {\em generic
%chaining} method~\cite{Tal05} which was originally developed as a
%tool for bounding similar quantities in the context of Gaussian
%processes (see also Talagrand's original treatment of this
%subject---a significant part of~\cite{Tal-original} is explicitly
%devoted to approximations of metrics by ultrametrics). Our idea to
%consider maximum gradient embeddings is partially motivated by
%Talagrand's work and our proof of the upper bound in
%Theorem~\ref{thm:ourmax} is a ``decomposition into annuli" argument
%which is reminiscent of Talagrand's approach.

%Talagrand's original method for bounding expectations of maxima of
%certain processes has been simplified and refined since its original
%appearance in~\cite{Tal-original}. But, it should be remarked here
%that a significant part of Talagrand's original
%paper~\cite{Tal-original} is explicitly devoted to approximations of
%metrics by ultrametrics. We believe that the similarity between this
%early appearance of ultrametric approximations and the more modern
%approaches deserves further investigation.

The heart of this paper is the lower bound in
Theorem~\ref{thm:ourmax}. The metrics $Y_n$ in
Theorem~\ref{thm:ourmax} are the diamond graphs of Newman and
Rabinovich~\cite{NR03}, which will be defined in
Section~\ref{section:diamond}. These graphs have been previously
used as counter-examples in several embedding problems---
see~\cites{GNRS04,NR03,BC05,LN04}. In particular, we were inspired to
consider these examples by the proof in~\cite{GNRS04} of the fact
that they require distortion $\Omega(\log n)$ in any probabilistic
embedding into trees. However, our proof of the $\Omega((\log n)^2)$
lower bound in Theorem~\ref{thm:ourmax} is considerably more
delicate than the proof in~\cite{GNRS04}. This proof, together with
other lower bounds for maximum gradient embeddings, is presented in
Section~\ref{section:diamond}.

\subsection{A framework for clustering problems with monotone
costs}\label{sec:framework}

We now turn to some algorithmic applications of
Theorem~\ref{thm:ourmax}. The general reduction in
Theorem~\ref{thm:general reduction} below should also be  viewed
as an explanation why maximum gradient embeddings are so
natural--- they are precisely the notion of embedding which allows
such reductions to go through.
%In Section~\ref{sec;axioms} we will
%present an even more general algorithmic framework, which is not
%limited to clustering problems, in which maximum gradient
%embeddings can be applied.
%As remarked above, we do not attempt to
%be encyclopedic here--- we will present a general paradigm for
%non-linear clustering problems, and analyze some examples. There
%are many more directions for future research that arise from the
%ideas described here. In particular, in a future paper we intend
%to apply maximum gradient embeddings to problems in online
%algorithms.

A general setting of the clustering problem is as follows. Let $X$
be an $n$-point set, and denote by $\MET(X)$ the set of all metrics
on $X$. A {\em possible clustering solution} consists of sets of the
form $\{(x_1,C_1),\ldots,(x_k,C_k)\}$ where $x_1,\ldots,x_k\in X$
and $C_1,\ldots,C_k\subseteq X$. We think of $C_1,\ldots,C_k$ as the
clusters, and $x_i$ as the ``center" of $C_i$. In this  general
framework we do not require that the clusters cover $X$, or that
they are pairwise disjoint, or that they contain their centers. Thus
the space of possible clustering solution is $\mathcal S\coloneqq
2^{X\times 2^X}$ (though the exact structure of $\mathcal S$ does
not play a role in the proof of Theorem~\ref{thm:general reduction}
below). Assume that for every point $x\in X$, every metric $d\in
\MET(X)$, and every possible clustering solution $P\in \mathcal S$,
we are given $\Gamma(x,d,P)\in [0,\infty]$, which we think of as a
measure of the dissatisfaction of $x$ with respect to $P$ and $d$.
Our goal is to minimize the average dissatisfaction of the points of
$X$. Formally, given a measure of dissatisfaction (which we also
call in what follows a {\em clustering cost function})
$\Gamma:X\times\MET(X)\times \mathcal S\to [0,\infty]$, we wish to
compute for a given metric $d\in \MET(X)$ the value
$$\Opt_\Gamma(X,d)\stackrel{\mathrm{def}}{=}\min\left\{\sum_{x\in X} \Gamma(x,d,P): P\in \mathcal S\right\}$$ (Since we are mainly concerned with the algorithmic
aspect of this problem, we assume from now on  that $\Gamma$ can be
computed efficiently.)

We make two natural assumptions on the cost function $\Gamma$. First
of all, we will assume that it scales homogeneously with respect to
the metric, i.e. for every $\lambda>0$, $x\in X$, $d\in \MET(X)$ and
$P\in \mathcal S$ we have $\Gamma(x,\lambda d,P)=\lambda
\Gamma(x,d,P)$. Secondly we will assume that $\Gamma$ is monotone
with respecting to the metric, i.e. if $d,\overline{d}\in \MET(X)$
and $x\in X$ satisfy $d(x,y)\le \overline{d}(x,y)$ for every $y\in
X$ then $\Gamma(x,d,P)\le \Gamma(x,\overline{d},P)$. In other words,
if all the points in $X$ are further with respect to $\overline{d}$
from $x$ then they are with respect to $d$, then $x$ is more
dissatisfied. This is a very natural assumption to make, as most
clustering problems look for clusters which are small in various
(metric) senses. We call clustering problems with $\Gamma$
satisfying these assumptions {\em monotone clustering problems}.
Essentially all the algorithmic minimization problems that have
benefitted from an application of~\eqref{eq:bartal} can be cast as
monotone clustering problems, but this framework also applies to
some ``non-linear" clustering optimization problems, as we shall see
presently.

The following theorem is a simple application of
Theorem~\ref{thm:ourmax}. It shows that it is enough to solve
monotone clustering problems on ultrametrics, with only a
polylogarithmic loss in the approximation factor.

\begin{theorem}[reduction to ultrametrics]\label{thm:general reduction} Let $X$ be an $n$-point set and
fix a homogeneous monotone clustering cost function
$\Gamma:X\times\MET(X)\times \mathcal S\to [0,\infty]$. Assume that
there is a randomized polynomial time algorithm which approximates $
\Opt_\Gamma(X,\rho)$ to within a factor $\alpha(n)$ on any
 ultrametric $\rho\in \MET(X)$. Then there is a randomized polynomial time
algorithm which approximates $\Opt_\Gamma(X,d)$ on any metric
$d\in \MET(X)$ to within a factor of $O\left(\alpha(n)(\log
n)^2\right)$.
\end{theorem}

\begin{proof}  Let $(X,d)$ be an $n$-point metric space
and let $\D$ be the distribution over random ultrametrics $\rho$
on $X$ from Theorem~\ref{thm:ourmax} (which is computable in
polynomial time, as follows directly from our proof of
Theorem~\ref{thm:ourmax} in Section~\ref{sec:proof theorem}). In
other words, $\rho(x,y)\ge d(x,y)$ for all $x,y\in X$ and
$$
\max_{x\in X}\  \E_\D \left[\max_{y\in X\setminus \{x\}}
\frac{\rho(x,y)}{d(x,y)}\right]\le C(\log n)^2.
$$
 Let
$P\in \mathcal S$ be a clustering solution for which
$$
\Opt_\Gamma(X,d)=\sum_{x\in X} \Gamma(x,d,P). $$
Using the
monotonicity and homogeneity of $\Gamma$ we see that

\begin{eqnarray*}
\Opt_\Gamma(X,\rho)&\le& \sum_{x\in X} \Gamma(x,\rho,P)\le
\sum_{x\in X} \Gamma\left(x,\left[\max_{y\in
X\setminus\{x\}}\frac{\rho(x,y)}{d(x,y)}\right]\cdot d,P\right)=
\sum_{x\in X} \left[\max_{y\in
X\setminus\{x\}}\frac{\rho(x,y)}{d(x,y)}\right]\cdot\Gamma(x,d,P).
\end{eqnarray*}
Taking expectation we conclude that
$$
\E_\D \left[\Opt_\Gamma(X,\rho)\right]\le \sum_{x\in X}
\left(\E_\D\left[\max_{y\in
X\setminus\{x\}}\frac{\rho(x,y)}{d(x,y)}\right]\right)\Gamma(x,d,P)\le
C(\log n)^2\cdot \Opt_\Gamma(X,d).
$$
Hence, with probability at least $\frac12$ we have
$$
\Opt_\Gamma(X,\rho)\le 2C(\log n)^2\cdot \Opt_\Gamma(X,d).
$$
For such $\rho$ compute a clustering solution $Q\in \mathcal S$
satisfying
$$
\sum_{x\in X} \Gamma(x,\rho,Q)\le \alpha(n)\Opt_\Gamma(X,\rho)\le
2C\alpha(n)(\log n)^2\cdot \Opt_\Gamma(X,d).
$$
Since $\rho\ge d$ it remains to use the monotonicity of $\Gamma$
once more to deduce that
\begin{eqnarray*}
\sum_{x\in X} \Gamma(x,\rho,Q)\ge \sum_{x\in X} \Gamma(x,d,Q)\ge
\Opt_\Gamma(X,d).
\end{eqnarray*}
Thus $Q$ is a $O\left(\alpha(n)(\log n)^2\right)$ approximate
solution to the clustering problem on $(X,d)$ with cost $\Gamma$.
\end{proof}

%Due to Theorem~\ref{thm:general reduction} we see that the main
%difficulty in monotone clustering problems  lies in the design of
%good approximation algorithms for them on ultrametrics.

Theorem~\ref{thm:general reduction} is a generic reduction, and in
many particular cases it might be possible use a case-specific
analysis to improve the $O\left((\log n)^2\right)$ loss in the
approximation factor. However, as a general reduction paradigm for
clustering problems, Theorem~\ref{thm:general reduction} makes it
clear why maximum gradient embeddings are natural.

We shall now demonstrate the applicability of the monotone
clustering framework to two concrete examples called {\em
fault-tolerant $k$-median clustering} and {\em $\Sigma \ell_p$
clustering}. We are not aware of a previous investigation of these
problems, but we believe that they are quite natural. It also seems
plausible that, just as in the problems for which Bartal's method
originally yielded the first non-trivial algorithmic results, a
better approximation factor might be obtainable via more
problem-specific tools.

%We will now present some specific applications of
%Theorem~\ref{thm:general reduction}. Before doing so we would like
%to state at the outset that we do not know if it is the case that
%all  monotone clustering problems can be well-approximated on
%ultrametrics. If this were true then Theorem~\ref{thm:general
%reduction} would yield an approximation algorithm for all
% monotone clustering problems. In
%Section~\ref{sec:dynamic} we will use a simple dynamic programming
%approach to design algorithms for certain monotone clustering
%problems. Our approach is quite flexible, and one can use it to give
%some additional conditions which show that a wide range of monotone
%clustering problems can be solved on ultrametrics. However carrying
%the analysis through in such generality seems to hide the simplicity
%of our ideas, so we prefer to analyze particular cases. In any case,
%we state the following natural question which remains open:

%\medskip

%\noindent{\bf Open problem:} Which monotone clustering problems
%have a polynomial time polylogarithmic approximation algorithm on
%ultrametrics?

%\medskip

%We now describe some monotone clustering problems for which
%Theorem~\ref{thm:general reduction} yields the best known
%approximation algorithm.

\bigskip

\noindent {\bf Fault-tolerant $k$-median and facility location.} The
$k$-median problem is as follows. Given an $n$-point metric space
$(X,d_X)$ and $k\in \mathbb N$, find $x_1,\ldots,x_k\in X$ that
minimize the objective function
\begin{eqnarray}\label{eq:def median} \sum_{x\in X} \min_{j\in \{x_1,\ldots,x_k\}} d_X(x,x_j).
\end{eqnarray}
This very natural and well studied problem can be easily cast as
monotone clustering problem by defining
$\Gamma(x,d,\{(x_1,C_1),\ldots,(x_m,C_m)\})$ to be $\infty$ if
$m\neq k$, and otherwise
$$\Gamma(x,d,\{(x_1,C_1),\ldots,(x_m,C_m)\})=\min_{j\in
\{x_1,\ldots,x_k\}} d(x,x_j).$$

The linear structure of~\eqref{eq:def median} makes it a prime
example of a problem which can be approximated using Bartal's
probabilistic embeddings. Indeed, the first non-trivial
approximation algorithm for $k$-median clustering was obtained by
Bartal in~\cite{Bartal98} (another such example is Min-Sum
clustering--- see~\cite{BCR01}). Since then this problem has been
investigated extensively: The first constant factor approximation
for it was obtained in~\cite{CGTS02} using LP rounding, and the
first combinatorial (primal-dual) constant-factor  algorithm was
obtained in~\cite{JV01}. In~\cite{AGKMMP04} an analysis of a
natural local search heuristic yields the best known approximation
factor for $k$-median clustering.

Here we study the following fault-tolerant version of the
$k$-median problem. Let $(X,d)$ be an $n$-point metric space and
fix $k\in \mathbb N$. Assume that for every $x\in X$ we are given
an integer $j(x)\in X$ (which we call the fault-tolerant parameter
of $x$). Given $x_1,\ldots,x_k$ and $x\in X$ let $x_j^*(x;d)$ be
the $j$-th closest point to $x$ in $\{x_1,\ldots,x_k\}$. In other
words, $\{x_j^*(x;d)\}_{j=1}^k$ is a re-ordering of
$\{x_j\}_{j=1}^k$ such that $ d(x,x_1^*(x;d))\le \cdots\le
d(x,x_k^*(x;d))$. Our goal is to minimize the objective function
\begin{eqnarray}\label{eq:def fault median} \sum_{x\in X} d\left(x,x_{j(x)}^*(x;d)\right).
\end{eqnarray}

 To understand~\eqref{eq:def fault median} assume for the sake of
simplicity that $j(x)=j$ for all $x\in X$. If $\{x_j\}_{j=1}^k$
minimize~\eqref{eq:def fault median} and $j-1$ of them are deleted
(due to possible noise), then we are still ensured that on average
every point in $X$ is close to one of the $x_j$. In this sense the
clustering problem in~\eqref{eq:def fault median} is fault-tolerant.
In other words, the optimum solution of~\eqref{eq:def fault median}
is insensitive to (controlled) noise. Observe that for $j=1$ we
return to the $k$-median clustering problem.

We remark that another fault-tolerant version of $k$-median
clustering was introduced in~\cite{JV04}. In this problem we
connect each point $x$ in the metric space $X$ to $j(x)$ centers,
but the objective function is the sum over $x\in X$ of the sum of
the distances from $x$ to all the $j(x)$ centers. Once again, the
linearity of the objective function seems to make the problem
easier, and in~\cite{SS03} a constant factor approximation is
achieved (this immediately implies that our version of
fault-tolerant $k$-median clustering, i.e. the minimization
of~\eqref{eq:def fault median}, has a $O\left(\max_{x\in X} j(x)
\right)$ approximation algorithm). In particular, the LP that was
previously used for $k$-median clustering naturally generalizes to
this setting. This is not the case for our fault-tolerant version
in~\eqref{eq:def fault median}. Moreover, the local search
techniques for $k$-median clustering (see for
example~\cite{AGKMMP04}) do not seem to be easily generalizable to
the case $j>1$, and in any case seem to require $n^{\Omega(j)}$
time, which is not polynomial even for moderate values of $j$.

Arguing as above in the case of $k$-median clustering we see that
the fault-tolerant $k$-median clustering problem in~\eqref{eq:def
fault median} is a monotone clustering problem. In
Section~\ref{sec:dynamic} we show that it can be solved exactly in
polynomial time on ultrametrics. Thus, in combination with
Theorem~\ref{thm:general reduction}, we obtain a $O\left((\log
n)^2 \right)$ approximation algorithm for the minimization
of~\eqref{eq:def fault median} on general metrics.

\begin{remark}
Facility location type problems have been studied extensively since
the 1960's--- we refer to the book~\cite{MF90}, and specifically to
the chapter on uncapacitated facility location~\cite{CNW90}, for a discussion of such problems. The
uncapacitated metric facility location problem is closely related to
$k$-median problem (indeed $k$-median can be reduced to it via
Lagrangian relaxation--- see~\cite{JV01}), and has been studied
extensively in recent years
(see~\cites{STA97,GK98,KPR00,JV01,JMSV03,CG05}). In the context
of~\eqref{eq:def fault median} we can also consider the following
fault-tolerant version of the facility location problem. Assume in
addition that we are given non-negative facility costs
$\{f_x\}_{x\in X}$. Then the goal is to minimize over all
$x_1,\ldots,x_k\in X$ the objective function
\begin{eqnarray}\label{eq:facility}
\sum_{j=1}^k f_{x_j} +\sum_{x\in X} d\left(x,x_{j(x)}^*(x;d)\right).
\end{eqnarray}
The case $j(x)\equiv 1$ reduces to the classical un-capacitated
metric facility location problem. The techniques presented here can
be easily generalized to yield a $O\left((\log n)^2 \right)$
approximation algorithm for the minimization of~\eqref{eq:facility}
as well.
\end{remark}

\bigskip

\noindent {\bf $\Sigma \ell_p$ clustering.} Another problem which
illustrates the usefulness of Theorem~\ref{thm:general reduction} is
the $\Sigma \ell_p$ clustering problem which we now describe. Our
argument for this problem is quite general, and it applies to more
cost functions, but it is beneficial to concentrate on a concrete
example. For $p\in [1,\infty]$ the $\Sigma \ell_p$ clustering
problem is as follows: For a metric space $(X,d)$ and $k\in \mathbb
N$ the goal is to find $x_1,\ldots,x_k\in X$ and a partition of $X$
into $k$ sets $C_1,\ldots,C_k\subseteq X$ which minimize the
objective function
\begin{eqnarray}\label{eq:lp cluster}
\sum_{j=1}^k \left(\sum_{x\in C_j} d(x,x_j)^p\right)^{1/p}.
\end{eqnarray}

When $p=1$ this becomes the $k$-median problem, and when $p=\infty$
this is the ``sum of the cluster radii" problem, which has been
studied in~\cite{CP04}. In both of these extreme cases there is a
constant factor approximation algorithm known, so we automatically
get a $O\left(\min\{n^{1/p},n^{1-1/p}\}\right)$ approximation
algorithm for~\eqref{eq:lp cluster}.  Here we shall use the
framework of Theorem~\ref{thm:general reduction} to give a
 $O\left((\log
n)^2\right)$ approximation algorithm for this problem for general
$p$.

Observe that the $\Sigma\ell_p$ clustering problems are monotone
clustering problems. Indeed, all we need to do is define
$\Gamma(x,d,\{(x_1,C_1),\ldots,(x_m,C_m)\})$ to be $\infty$ if
$\{C_1,\ldots,C_m\}$ is not a partition of $X$ or $m\neq k$.
Otherwise set $\Gamma(x,d,\{(x_1,C_1),\ldots,(x_k,C_k)\})=0$ if
$x\notin \{x_1,\ldots,x_k\}$ and for $j\in \{1,\ldots,k\}$,
\begin{eqnarray*}
\Gamma(x_j,d,\{(x_1,C_1),\ldots,(x_k,C_k)\})=\left(\sum_{x\in C_j}
d(x,x_j)^p\right)^{1/p}.
\end{eqnarray*}
This definition clearly makes $\Gamma$ a homogeneous monotone
clustering cost function for any $p\in [1,\infty]$. The following
lemma, combined with Theorem~\ref{thm:general reduction}, therefore
implies that the $\Sigma \ell_p$ clustering problem has a
$O\left((\log n)^2\right)$ approximation algorithm.

\begin{lemma}\label{lem:PTAS} The $\Sigma\ell_p$ clustering
problem has a constant factor polynomial time approximation
algorithm (even a FPTAS) on ultrametrics.
\end{lemma}

Lemma~\ref{lem:PTAS} will be proved via dynamic programming in
Section~\ref{sec:dynamic}.

\section{Proof of the upper bound in Theorem~\ref{thm:ourmax}}\label{sec:proof
theorem}

We start by recalling some terminology and results concerning
random partitions of metric spaces. Given a partition $\mathscr P$
of a finite metric space $(X,d_X)$ and $x\in X$ we denote by
$\mathscr P(x)$ the unique element of $\mathscr P$ to which $x$
belongs. For $\Delta>0$ the partition $\mathscr P$ is said to be
$\Delta$-bounded if for every $x\in X$ we have $\diam(\mathscr
P(x))\le \Delta$. We also fix a positive measure $\mu$ on $X$. The
following fundamental result is due to~\cite{FRT03} when $\mu$ is
the uniform measure on $X$. The case of general measures was
observed in~\cites{LN05,KLMN05}, and the specific numerical
constants used below are taken from~\cite{MN05}.

\begin{lemma}\label{lem:FRT}
For every $\Delta>0$ there exists a distribution over
$\Delta$-bounded partitions $\mathscr P$ of $X$ such that for
every $x\in X$ and every $0<t\le \Delta/8$,
\begin{eqnarray}\label{eq:FRT}
\Pr\left[B_X(x,t)\not \subseteq \mathscr P(x) \right]\le
\frac{16t}{\Delta}\cdot\log\frac{\mu(B_X(x,\Delta))}{\mu(B_X(x,\Delta/8))}.
\end{eqnarray}
\end{lemma}

We also recall the notion of a {\em quotient of a metric space}
(see~\cites{MN04,Gromov99,BH99}). Let $\W=\{W_1,\ldots,W_m\}$ be a
partition of $X$. For $W,W'\in \W$ write
$d_X(W,W')=\min\{d_X(x,y):\ x\in W,\ y\in W'\}$. The quotient
metric space $(X/\W,d_{X/\W})$ is define as follows. As a set
$X/\W$ coincides with $\W$. The metric $d_{X/\W}$ is the maximal
metric on $\W$ which is majorized by $d_X(\cdot,\cdot)$. In other
words, for $W,W'\in \W$,
$$
d_{X/\W}(W,W')=\min \left\{\sum_{j=1}^{m-1}d_X(V_{j-1},V_j):\
V_0,\ldots,V_{m-1}\in \W,\ V_0=W, \ V_{m-1}=W' \right\}.
$$
Note that the $V_j$'s in the definition above need not be distinct.

The following lemma is a well known ``quotient version" of
Lemma~\ref{lem:FRT}. The argument dates back at least to
Bartal~\cite{Bartal96}, and appeared in various guises in several
other places--- see for example~\cites{HM05,MN05}. Since we couldn't
locate the formulation that we need in the literature, we include a
proof here.

\begin{lemma}\label{lem:quotient}
Let $(X,d_X)$ be an $n$-point metric space and $\Delta>0$. Then
there exists a distribution over $\Delta$-bounded partitions
$\mathscr P$ of $X$ such that for every $x,y\in X$, if $d_X(x,y)\le
\frac{\Delta}{2n}$ then $\mathscr P(x)=\mathscr P(y)$, and for every
$x\in X$ and $0<t\le \Delta/16$,
$$
\Pr\left[B_X(x,t)\not\subseteq \mathscr
P(x)\right]\le\frac{32t}{\Delta}\cdot \log
\frac{\mu(B_X(x,\Delta))}{\mu(B_X(x,\Delta/16))}.
$$
\end{lemma}

\begin{proof} Define an equivalence relation on $X$ by $x\sim y$ if
there exists $k\in \N$ and $x_0,\ldots,x_k\in X$ such that $x_0=x$,
$x_k=y$ and $d_X(x_{i-1},x_i)\le \frac{\delta}{2n}$ for all $i\in
\{1,\ldots,k\}$.
%Let $\rho$ be an ultrametric on $X$ such that
% $d_X(x,y)\le\rho(x,y)\le nd_X(x,y)$ for every $x,y\in X$. The simple
%construction of such a metric is contained in Lemma 3.6
%in~\cite{BLMN05}. Define an equivalence relation on $X$ by $x\sim
%y$ if $\rho(x,y)\le \frac{\Delta}{2n}$. This is an equivalence
%relation since $\rho$ is an ultrametric.
Let $\mathscr W=\{W_1,\ldots,W_m\}$ be the equivalence classes of
this relation, and consider the quotient metric space $X/\mathscr
W$. We also denote by $\pi:X\to \W$ the induced quotient map, i.e.
for $x\in W_j$, $\pi(x)=W_j$. Let $\mu\circ \pi^{-1}$ be the measure
on $\W$ given for $W\in \W$ by $\mu\circ
\pi^{-1}(W)=\mu(\pi^{-1}(W))$. Observe that for every $x,y\in X$,
\begin{eqnarray}\label{eq:equiv quotient}
d_X(x,y)-\frac{\Delta}{2}\le d_{X/\W}(\pi(x),\pi(y))\le d_X(x,y).
\end{eqnarray}
Indeed, the upper bound in~\eqref{eq:equiv quotient} is immediate
from the definition of a quotient metric. The lower bound
in~\eqref{eq:equiv quotient} is proved as follows. There are points
$x=x_0,x_1,\ldots,x_{m-1}=y$ in $X$ such that the sets $\{\pi(x_j)\}_{j=0}^{m-1}$ are distinct (and hence disjoint), and
$d_{X/\W}(\pi(x),\pi(y))=\sum_{j=1}^{m-1}
d_X(\pi(x_{j-1}),\pi(x_j))$. For $j\in \{1,\ldots,m-1\}$ let $a_j\in
\pi(x_{j-1})$ and $b_j\in \pi(x_j)$ be such that
$d_X(a_j,b_j)=d_X(\pi(x_{j-1}),\pi(x_j))$. Since, by the definition
of the equivalence relation $\sim$, for all $z\in X$ we have
%beginCHANGE 
$\diam(\pi(z))=\max_{a,b\in\pi(z)}d_X(a,b)\le \frac{(|\pi(z)|-1) \Delta}{2n}$ we get that
\begin{multline*}
d_X(x,y)
\le d_X(x,a_1)+ \sum_{j=1}^{m-1}d_X(a_j,b_j)+ \sum_{j=1}^{m-2}d_X(b_j,a_{j+1})+ d_X(b_{m-1},y) \\
\le \sum_{j=0}^{m-1} \frac{(|\pi(x_j)|-1)\Delta}{2n} + d_{X/\mathscr{W}}(\pi(x),\pi(y))
\le \frac{\Delta}{2} + d_{X/\mathscr{W}}(\pi(x),\pi(y)),
\end{multline*}
%endCHANGE 
implying the lower bound in~\eqref{eq:equiv quotient}.

 Let $\mathscr Q$ be a distribution
over $\Delta/2$-bounded partitions of $X/\W$ such that for every
$W\in \W$ and every $0<t\le\Delta/16$ we have
\begin{eqnarray}\label{eq:in the quotient}
\Pr\left[B_{X/\W}(W,t)\not\subseteq \mathscr Q(W)\right]\le
\frac{32t}{\Delta}\cdot\log\frac{\mu\circ\pi^{-1}(B_{X/\W}(W,\Delta/2))}{\mu\circ
\pi^{-1} ( B_{X/\W}(W,\Delta/16))}.
\end{eqnarray}
The existence of $\mathscr Q$ follows from Lemma~\ref{lem:FRT}. Let
$\mathscr P$ be the partition of $X$ given by $\mathscr
P=\{\pi^{-1}(A):\ A\in \mathscr Q\}$. Note that~\eqref{eq:equiv
quotient} implies that for every $x\in X$ we have
$\pi^{-1}\left(B_{X/\W}(\pi(x),\Delta/2)\right)\subseteq
B_X(x,\Delta)$ and for every $t>0$,
$\pi^{-1}\left(B_{X/\W}(\pi(x),t)\right)\supseteq B_X(x,t)$.
Thus~\eqref{eq:in the quotient} implies that for every $x\in X$ and
$0<t\le \Delta/16$,
$$
\Pr\left[B_X(x,t)\not\subseteq \mathscr P(x)\right]\le
\Pr\left[B_{X/\W}(\pi(x),t)\not\subseteq \mathscr
Q(\pi(x))\right]\le\frac{32t}{\Delta}\cdot \log
\frac{\mu(B_X(x,\Delta))}{\mu(B_X(x,\Delta/16))}.
$$
It remains to note that~\eqref{eq:equiv quotient} implies that
$\mathscr P$ is $\Delta$-bounded and if $d_X(x,y)\le
\frac{\Delta}{2n}$ then $x\sim y$, which means that $\pi(x)=\pi(y)$,
so that $\mathscr P(x)=\mathscr P(y)$.
\end{proof}

\begin{proof}[Proof of the upper bound in Theorem~\ref{thm:ourmax}] For every $k\in \Z$  let $\mathscr P_k$
be a random partition sampled from
 the distribution over partitions of $X$ from Lemma~\ref{lem:quotient} with $\Delta=16^k$, where $\mu$ is the counting measure on $X$ (we assume in
what follows that the distributions for different values of $k$ are
independent). For $x,y\in X$ let $k$ be the largest integer for
which $\mathscr P_k(x)\neq \mathscr P_k(y)$ (such a $k$ must exists
since  for small enough $k$ we have $\mathscr P_k(z)=\{z\}$ for all
$z\in X$). Denote $\rho(x,y)=16^{k+1}$. Then $\rho$ is a (random)
ultrametric on $X$. Indeed, if $x,y,z\in X$ and $\rho(x,y)=16^{k+1}$
then $\mathscr P_k(x)\neq \mathscr P_k(y)$. It follows that either
$\mathscr P_k(z)\neq \mathscr P_k(x)$ or $\mathscr P_k(z)\neq
\mathscr P_k(y)$. Thus by the definition of $\rho$ we have that
$\max\{\rho(x,z),\rho(y,z)\}\ge \rho(x,y)$. Note also that if
$\rho(x,y)=16^{k+1}$ then $\mathscr P_{k+1}(x)= \mathscr
P_{k+1}(y)$, so that $d_X(x,y)\le \diam(\mathscr P(x))\le
16^{k+1}=\rho(x,y)$. It follows that the identity mapping on $X$ is
a random non-contractive embedding of $X$ into the ultrametric
$(X,\rho)$. Finally, since whenever $d_X(x,y)\le \frac{16^k}{2n}$ we
have $\mathscr P_k(x)= \mathscr P_k(y)$, we are ensured that
$\rho(x,y)\le 32 nd_X(x,y)$ for every $x,y\in X$.

Denote for $x\in X$ and $i\in \Z$, $A_i(x)=B_X(x,16^{i})\setminus
B_X(x,16^{i-1})$. For every $j\in \mathbb N$ and $k\in \mathbb Z$ if
$B_X(x,16^{k-j})\subseteq \mathscr P_k(x)$ then for every $y\in
B_X(x,16^{k-j})$ we have $\mathscr P_k(x)=\mathscr P_k(y)$, and
therefore by the definition of $\rho(x,y)$ we have $\rho(x,y)\le
16^k$. Thus, if $y\in A_{k-j}(x)$ we have $\rho(x,y)\le
16^k<16^{j+1}d_X(x,y)$. This establishes the following inclusion of
events:
\begin{eqnarray*}\label{eq:inclusion}
\left\{\max_{y\in A_{k-j}(x)}\frac{\rho(x,y)}{d_X(x,y)}\ge
16^{j+1}\right\}\subseteq \left\{B_X(x,16^{k-j})\not\subseteq
\mathscr P_k(x) \right\}.
\end{eqnarray*}
%Indeed, assume for the sake of contradiction
%that~\eqref{eq:inclusion} fails. Then it is possible that
%$B_X(x,16^{k-j})\subseteq \mathscr P_k(x)$, yet there exists $y\in
%A_{k-j}(x)$ satisfying $16^{j+2}d_X(x,y)>  \rho(x,y)\ge
%16^{j+1}d_X(x,y)$. Let $s\in \mathbb N$ be such that
%$\rho(x,y)=16^{s+1}$. Since $y\in B_X(x,16^{k-j})$ we know that
%$16^{s+1}=\rho(x,y)<16^{j+2}\cdot 16^{k-j}=16^{k+2}$. Thus $s\le k$.
%Since $y\in B_X(x,16^{k-j})\subseteq \mathscr P_k(x)$ we also know
%that $\mathscr P_k(x)=\mathscr P_k(y)$, which by the definition of
%$\rho$ implies that $s\neq k$. Thus $s\le k-1$. But since $y\notin
%B_X(x,16^{k-j-1})$ we have that $\rho(x,y)=16^{s+1}\le
%16^{k}<16^{j+1}d_X(x,y)$, which is a contradiction to our assumption
%on $y$.
%Using~\eqref{eq:inclusion} we have
hence
\begin{eqnarray*}
\Pr\left[\max_{y\in A_{k-j}(x)} \frac{\rho(x,y)}{d_X(x,y)}\ge
16^{j+1} \right]\le \Pr\left[B_X(x,16^{k-j})\not\subseteq \mathscr
P_k(x) \right]\le \frac{32}{16^j}\cdot \log
\frac{|B_X(x,16^k)|}{|B_X(x,16^{k-1})|}.
\end{eqnarray*}
Thus, since $X=\bigcup_{i\in \Z} A_i(x)$, we see that
\begin{multline}\label{eq:weak}
\Pr\left[\max_{y\in X\setminus\{x\}}\frac{\rho(x,y)}{d_X(x,y)}\ge
16^j \right]= \Pr\left[\bigcup_{i\in \Z}\left\{ \max_{y\in
A_{i}(x)}\frac{\rho(x,y)}{d_X(x,y)}\ge 16^j \right\}\right] \le
\sum_{i\in \Z} \Pr\left[\max_{y\in A_i(x)}
\frac{\rho(x,y)}{d_X(x,y)}\ge 16^j \right]\\ \le \sum_{i\in \Z}
\frac{32}{16^{j-1}}\cdot \log
\frac{|B_X(x,16^{i+j-1})|}{|B_X(x,16^{i+j-2})|} \le
\frac{512}{16^j}\cdot\log n.
\end{multline}
It follows that there exists a universal constant $C>0$ such that
for all $u>0$ we have
$$
\Pr\left[\max_{y\in X\setminus\{x\}}\frac{\rho(x,y)}{d_X(x,y)}\ge u
\right]\le \frac{C\log n}{u}.
$$
Hence, using the a priori bound $\rho(x,y)\le 32 nd_X(x,y)$, it
follows that
$$
\E\left[\max_{y\in
X\setminus\{x\}}\frac{\rho(x,y)}{d_X(x,y)}\right]=\int_0^{32
n}\Pr\left[\max_{y\in X\setminus\{x\}}\frac{\rho(x,y)}{d_X(x,y)}\ge
u \right]du\le \int_0^{32n} \min\left\{1,\frac{C\log
n}{u}\right\}du=O\left(1+(\log n)^2\right).
$$
%\begin{multline*}
%\E \left[\max_{y\in
%X\setminus\{x\}}\frac{\rho(x,y)}{d_X(x,y)}\right]\le
%16\Pr\left[\max_{y\in X\setminus\{x\}}\frac{\rho(x,y)}{d_X(x,y)}\le
%16\right]+ \sum_{j=1}^{\lfloor \log_{16}
%(32n^2)\rfloor}16^{j+1}\Pr\left[16^{j+1}>\max_{y\in
%X\setminus\{x\}}\frac{\rho(x,y)}{d_X(x,y)}\ge 16^j \right]\\
%\le 16+\sum_{j=1}^{\lfloor \log_{16} (32n^2)\rfloor}16^{j+1}\cdot
%\frac{512}{16^j}\cdot\log n = O\left(1+(\log n)^2\right).
%\end{multline*}
This completes the proof of the upper bound in
Theorem~\ref{thm:ourmax}.
\end{proof}

\begin{remark} The above argument also shows that for every
$n$-point metric space $(X,d_X)$ there exists a distribution over
non-contractive embeddings into ultrametrics $f:X\to U$ such that
$$
\E_\D \left[|\nabla f(x)|_\infty\right]=O\left(1+(\log n)\log \Phi(X)\right),
$$
where $\Phi(X)$ is the aspect ratio of $X$, which is defined by
$$
\Phi(X)=\frac{\diam{X}}{\min_{\substack{x,y\in X\\x\neq
y}}d_X(x,y)}=\frac{\max_{x,y\in X}d_X(x,y)}{\min_{\substack{x,y\in
X\\x\neq y}}d_X(x,y)}.
$$
\end{remark}

\section{Tight lower bounds for cycles, paths, and diamond
graphs}\label{section:diamond}

As mentioned in the introduction, the metrics $Y_n$ in
Theorem~\ref{thm:ourmax} are the diamond graphs of Newman and
Rabinovich~\cite{NR03}, which will be defined presently. Before
passing to this more complicated (and strongest) lower bound, we
will analyze the simpler examples of cycles and paths, which are
of independent interest.

Let $C_n$, $n>3$, be the unweighted path on $n$-vertices. We will
identify $C_n$ with the group $\Z_n$ of integers modulo $n$. We
first observe that in this special case the upper bound in
Theorem~\ref{thm:ourmax} can be improved to $O(\log n)$. This is
achieved by using Karp's embedding of the cycle into spanning
paths--- we simply choose an edge of $C_n$ uniformly at random and
delete it. Let $f:C_n\to \Z$ be the randomized embedding thus
obtained, which is clearly non-contractive.

As Karp observed, one can readily verify that as a probabilistic
embedding into trees $f$ has distortion at most $2$. We will now
show that as a maximum gradient embedding, $f$ has distortion
$\Theta(\log n)$. Indeed, fix $x\in C_n$, and denote the deleted
edge by $\{a,a+1\}$. Assume that $d_{C_n}(x,a)=t\le n/2-1$. Then the
distance from $a+1$ to $x$ changed from $t+1$ in $C_n$ to $n-t-1$ in
the path. It is also easy to see that this is where the maximum
gradient is attained. Thus
$$
\E \left[|\nabla f(x)|_\infty \right]\approx \frac{2}{n}\sum_{0\le t\le
n/2}\frac{n-t-1}{t+1}=\Theta(\log n).
$$
We will now show that any maximum gradient embedding of $C_n$ into
a distribution over trees incurs distortion $\Omega(\log n)$. For
this purpose we will use the following  lemma from \cite{RR98}.
\begin{lemma} \label{lem:rr}
For any tree metric $T$, and any non-contractive embedding $g:C_n
\to T$, there exists an edge $(x,x+1)$ of $C_n$ such that
$d_T(g(x),g(x+1))\geq \frac{n}{3}-1$.
\end{lemma}

Now, let $\D$ be a distribution over non-contractive embeddings of
$C_n$ into trees $f:C_n\to T$. By Lemma~\ref{lem:rr} we know that
there exists $x\in C_n$ such that $d_T(f(x),f(x+1))\ge
\frac{n-3}{3}$. Thus for every $y\in C_n$ we have that
$\max\{d_T(f(y),f(x)),d_T(f(y),f(x+1))\}\ge \frac{n-3}{6}$. On the
other hand $\max\{d_{C_n}(y,x),d_{C_n}(y,x+1)\}\le
d_{C_n}(x,y)+1$. It follows that
$$
|\nabla f(y)|_\infty\ge \frac{n-3}{6d_{C_n}(x,y)+6}.
$$
Summing this inequality over $y\in C_n$ we see that
$$
\sum_{y\in C_n}|\nabla f(y)|_\infty\ge \sum_{0\le k\le
n/2}\frac{n-3}{6k+6}=\Omega(n\log n).
$$
Thus
$$
\max_{y\in C_n} \E_\D \left[|\nabla f(y)|_\infty\right]\ge
\frac{1}{n}\sum_{y\in C_n}\E_\D|\nabla f(y)|_\infty=\Omega(\log
n),
$$
as required.

\medskip

We will now deal with the more complicated case of maximum
gradient embeddings of the unweighted path on $n$-vertices, which
we denote by $P_n$, into ultrametrics. The following proposition
shows that Theorem~\ref{thm:ourmax} is optimal when one considers
embeddings into ultrametrics. This is weaker than the lower bound
in Theorem~\ref{thm:ourmax}, which deals with embeddings into
arbitrary trees (note that $P_n$ is a tree).

\begin{proposition}\label{prop:path} Let $\D$ be a distribution
over non-contractive embeddings of $P_n$ into ultrametrics
$f:P_n\to U$. Then there exists $x\in P_n$ such that $\E_\D\left[
|\nabla f(x)|_\infty\right] =\Omega((\log n)^2)$.
\end{proposition}

Before proving Proposition~\ref{prop:path} we record the following
numerical inequalities.

\begin{lemma}\label{lem:numerical} The following elementary
inequalities hold true:
\begin{enumerate}
\item For every $a,b\in \{0,1,2,\ldots\}$,
$$
a(\log a)^2+b(\log b)^2\ge
(a+b)\left(\log(a+b)\right)^2-2\left[1+\log\left(\frac{a+b}{a}\right)\right]a\log(a+b).
$$
\item For every $x\ge 1$, $
 \left(1+\log x\right)\log x\le 4\sqrt x$.
\end{enumerate}
\end{lemma}

\begin{proof} The first inequality is trivial if $a=0$ or
$b=0$, so assume that $a,b\ge 1$. Denote for $t\ge 0$,
$\psi(t)=t(\log t)^2$. Then
\begin{eqnarray*}
(a+b)\left(\log(a+b)\right)^2-b(\log b)^2&=&\int_b^{a+b}
\psi'(t)dt\\
&=& \int_b^{a+b}\left[(\log t)^2+2\log t\right]dt\\
&\le& a\left(\log(a+b)\right)^2+2a\log(a+b)\\
&=& a(\log a)^2+a\left[\log(a+b)+\log a\right]\cdot
\log\left(\frac{a+b}{a}\right)+2a\log(a+b)\\
&\le&  a(\log a)^2+
2\left[1+\log\left(\frac{a+b}{a}\right)\right]a\log(a+b),
\end{eqnarray*}
proving the first assertion in Lemma~\ref{lem:numerical}.

The second assertion in Claim~\ref{lem:numerical} follows from the
inequality $\log x\le 2\sqrt[4]{x}-1$, which is true since the
minimum of the function $y\mapsto 2\sqrt[4]{y}-1-\log y$, which is
attained at $y=16$, is positive.
\end{proof}

\begin{proof}[Proof of Proposition~\ref{prop:path}] We think of $P_n$ as the interval of integers  $I=\{0,\ldots,n-1\}\subseteq \R$.
Arguing the same as in the case of the cycle $C_n$, it is enough
to prove that if $(U,d_U)$ is an ultrametric and $f:P_n\to U$ is
non-contractive then
\begin{eqnarray}\label{eq:poincare goal path}
\frac{1}{n}\sum_{x=0}^{n-1}|\nabla f(x)|_\infty \ge c(\log n)^2,
\end{eqnarray}
where $c>0$ is a universal constant.

Given a sub-interval $J=\{a,a+1,\ldots,a+t\}\subseteq
\{0,\ldots,n-1\}$ let $m_J$ be the largest point $m\in
\{a+1,\ldots,a+t\}$ for which
$d_U(f(m-1),f(m))=\|f|_J\|_{\Lip}=\max_{1\le i\le t}
d_U(f(a+i-1),f(a+i))$ (if $t=0$ then we set $m_J=a$). Since the
distortion of $J$ in any embedding into an ultrametric is at least
$|J|-1$ (see Lemma 2.4 in~\cite{MN04}), we know that
$d_U(f(m_J-1),f(m_J))\ge t=|J|-1$. We shall denote in what follows
$J_s$ to be the shorter of the two intervals
$\{a,a+1,\ldots,m_J-1\}$ and $\{m_J,\ldots,a+t\}$ (breaking ties
arbitrarily), and $J_b$ will denote the longer of these two
intervals (when $|J|=1$ we use the convention $J_s=J_b$). Thus
$J=J_s\cup J_b$ and $|J_s|\le |J_b|$. Finally, let $x_J$ be the
point in $J_s$ which is closest to $J_b$ (so that $x_J\in
\{m_J,m_{J-1}\}$).

We define a function $g_J:J\to \R$ inductively as follows. If $1\le
|J_s|\le \sqrt{|J|}$ then
\begin{equation}\label{eq:cases1}
g_J(x)=\begin{cases} g_{J_s}(x) \ & \text{if }  x\in J_s\setminus\{x_J\},\\
\frac18\left[1+\log\left(\frac{|J|}{|J_s|}\right)\right]|J_s|\log|J| \ & \text{if } x=x_J,\\
g_{J_b}(x) \ & \text{if } x\in J_b.
\end{cases}
\end{equation}
If, on the other hand, $|J_s|>\sqrt{|J|}$ then
\begin{equation}\label{eq:cases2}
g_J(x)=\begin{cases} g_{J_s}(x) \ & \text{if } x\in J_s \text{ and }|x-x_J|>\sqrt[4]{|J_s|},\\
\tfrac{|J|-1}{|x-x_J|+1} \ & \text{if } x\in J_s \text{ and } |x-x_J|\le\sqrt[4]{|J_s|},\\
g_{J_b}(x) \ & \text{if } x\in J_b.
\end{cases}
\end{equation}

The following claim summarizes the crucial properties of the these
mappings. Recall that we are using the notation
$I=\{0,\ldots,n-1\}$.

\begin{claim}\label{claim:properties g} The following assertions
hold true for every sub-interval $J\subseteq I$.
\begin{enumerate}
\item For every $x\in J$ we have $g_J(x)\le |\nabla (f|_J)(x)|_\infty=\max_{y\in J\setminus
\{x\}}\frac{d_U(f(x),f(y))}{|x-y|}$.
\item For every $x\in J$, $g_J(x)\le |J|-1$.
\item If $|J_s|\ge \sqrt{J}$ and $|x-x_J|\le\sqrt[4]{|J_s|}$
then $g_{J_s}(x)\le 4\sqrt{|J_s|}$.
\end{enumerate}
\end{claim}

\begin{proof} The proofs of all of the assertions in
Claim~\ref{claim:properties g} will be by induction on $J$. To
prove the first assertion assume first that $1\le |J_s|\le
\sqrt{|J|}$. From the recursive definition in~\eqref{eq:cases1} it
follows that we should show that
$\frac18\left[1+\log\left(\frac{|J|}{|J_s|}\right)\right]|J_s|\log|J|\le
|\nabla (f|_J)(x_J)|_\infty$. Since $x_J\in \{m_J-1,m_J\}$ the
definition of $m_J$ implies that $|\nabla (f|_J)(x_J)|\infty\ge
|J|-1$. Thus it is enough to show that
$\frac18\left(1+\log|J|\right)\sqrt{|J|}\log|J|\le |J|-1$, which
follows from the second assertion in Lemma~\ref{lem:numerical}.
If, on the other hand, $|J_s|> \sqrt {|J|}$ then from the
recursive definition in~\eqref{eq:cases2} it follows that it is
enough to show that for every $x\in J_s$ we have
$\frac{|J|-1}{|x-x_J|+1}\le |\nabla (f|_J)(x)|_\infty$. But since
$U$ is an ultrametric we know that
$$
|J|-1\le d_U(f(m_{J}-1),f(m_J))\le
\max\{d_U(f(x),f(m_{J}-1)),d_U(f(x),f(m_J))\},
$$ which
implies the required lower bound  on $|\nabla (f|_J)(x)|_\infty$
since $x_J\in \{m_J-1,m_J\}$. The second assertion in
Claim~\ref{claim:properties g} is proved similarly.

It remains to prove the third assertion in
Lemma~\ref{claim:properties g}. Let $K\subseteq J_s$ be the
sub-interval of $J_s$ in which the value of $g_{J_s}(x)$ was first
set. In other words, $K\subseteq J_s$ is the smallest interval for
which $x\in K_s$ and $g_{K}(x)= g_{J_s}(x)$. It follows in
particular that $|x-x_K|\le \sqrt[4]{|K_s|}$. Also, by
construction it is always the case that either $K_s$ or $K_b$ is
contained in the interval $[\min\{x_K,x_J\},\max\{x_K,x_J\}]$.
Since $K_s$ is shorter than $K_b$ we are assured that
\begin{eqnarray}\label{eq:bound length}
|K_s|\le |x_K-x_J|\le |x_K-x|+|x-x_J|\le \sqrt[4]{|K_s|}+
\sqrt[4]{|J_s|}\le  2\sqrt[4]{|J_s|}.
\end{eqnarray}

If $|K_s|\le \sqrt{|K|}$ then necessarily $x=x_K$ and $g_K(x)$ was
determined by the second line in~\eqref{eq:cases1}. Hence
\begin{eqnarray}\label{eq:3/4}
g_{J_s}(x)=g_K(x)=
\frac18\left[1+\log\left(\frac{|K|}{|K_s|}\right)\right]|K_s|\log|K|\le
\frac14\left[1+\log |J_s|\right]\sqrt[4]{|J_s|}\log |J_s|\le
4\sqrt{|J_s|},
\end{eqnarray}
where we used~\eqref{eq:bound length} and the last inequality
in~\eqref{eq:3/4} follows from the second assertion of
Lemma~\ref{lem:numerical}.

Otherwise $|K_s|> \sqrt{|K|}$ and $g_K(x)$ was determined by the
second line in~\eqref{eq:cases2}, i.e.
$$
g_{J_s}(x)=g_K(x)=\frac{|K|-1}{|x-x_K|+1}<|K|<|K_s|^2\le
4\sqrt{|J_s|},
$$
where we used~\eqref{eq:bound length}. This completes the proof of
Claim~\ref{claim:properties g}.
\end{proof}

With Claim~\ref{claim:properties g} at hand we are in position to
conclude the proof of Proposition~\ref{prop:path}. We will prove by
induction on $|J|$ that
\begin{eqnarray}\label{eq:inductive J}
\sum_{x\in J} g_J(x)\ge c |J|(\log |J|)^2.
\end{eqnarray}
This will prove~\eqref{eq:poincare goal path}, and hence imply
Proposition~\ref{prop:path}, since by the first assertion of
Claim~\ref{claim:properties g} we get that
$$
\sum_{x=0}^{n-1}|\nabla f(x)|_\infty\ge \sum_{x\in I} g_I(x)\ge c
n(\log n)^2.
$$

Inequality~\eqref{eq:inductive J} trivially holds true with small
enough constant $c$ if $|J|\le 2^{60}$, so assume that
$|J|>2^{60}$. To prove~\eqref{eq:inductive J} we distinguish
between two cases. If $|J_s|\le \sqrt{|J|}$ then since
$g_{J_s}(x_J)\le |J_s|$ (by the second assertion in
Claim~\ref{claim:properties g}) we see by induction that
\begin{eqnarray}
\!\!\!\!\!\!\!\!\!\sum_{x\in J} g_J(x)&=&\sum_{x\in J_s}
g_{J_s}(x)+\sum_{x\in J_b}
g_{J_b}(x) +g_J(x_{J})-g_{J_s}(x_J)      \nonumber\\
&>& c\left(|J_s|(\log |J_s|)^2+|J_b|(\log
|J_b|)^2\right)+2\left[1+\log\left(\frac{|J|}{|J_s|}\right)\right]|J_s|\log|J|-|J_s|\label{eq:induction}\\
&\ge& c|J|(\log
|J|)^2-2c\left[1+\log\left(\frac{|J|}{|J_s|}\right)\right]|J_s|\log|J|+\left[1+\log\left(\frac{|J|}{|J_s|}\right)\right]|J_s|\log|J|\label{eq:use numerical}\\
&\ge& c|J|(\log |J|)^2\label{eq:c small},
\end{eqnarray}
where in~\eqref{eq:induction} we used the inductive hypothesis and
the inductive definition in~\eqref{eq:cases1} , in~\eqref{eq:use
numerical} we used Lemma~\ref{lem:numerical}, and~\eqref{eq:c
small} holds for $c\le \frac12$.

On the other hand if $|J_s|> \sqrt{|J|}$ then
\begin{eqnarray}
\sum_{x\in J} g_J(x)&=&\sum_{x\in J_s}
g_{J_s}(x)+\sum_{x\in J_b} g_{J_b}(x)+\sum_{\substack{x\in J_s\\
|x-x_J|\le
\sqrt[4]{|J_s|}}}\left(\frac{|J|-1}{|x-x_J|+1}-g_{J_s}(x)\right)\label{eq:use the cases2}\\
&\ge& c|J|(\log
|J|)^2-2c\left[1+\log\left(\frac{|J|}{|J_s|}\right)\right]|J_s|\log|J|+\sum_{k=0}^{\left\lfloor
\sqrt[4]{|J_s|}
\right\rfloor}\frac{|J|-1}{k+1}-8|J_s|^{3/4}\label{eq:use the
claim}\\\nonumber
&\ge&c|J|(\log|J|)^2-2c\left[1+\log\left(\frac{|J|}{|J_s|}\right)\right]|J_s|\log|J|+\frac14(|J|-1)\log|J_s|-8|J|^{3/4}\label{eq:hold for huge J}\\
&\ge&c|J|(\log|J|)^2-2c\left[1+\log\left(\frac{|J|}{|J_s|}\right)\right]|J_s|\log|J|+\frac18(|J|-1)\log|J_s|\\
&\ge& c|J|(\log|J|)^2,\label{eq:hold for tiny c}
\end{eqnarray}
where in~\eqref{eq:use the cases2} we used the inductive
definition in~\eqref{eq:cases2}, in~\eqref{eq:use the claim} we
used the inductive hypothesis, Lemma~\ref{lem:numerical} and
Claim~\ref{claim:properties g}, and inequalities~\eqref{eq:hold
for huge J} and~\eqref{eq:hold for tiny c} hold for $|J|>2^{60}$
and small enough $c$, respectively, since $\frac{|J|}{2}\le |J_s|>
\sqrt{|J|}$. This completes the proof of
Proposition~\ref{prop:path}.
\end{proof}

We now pass to the proof of the lower bound in
Theorem~\ref{thm:ourmax} in its full strength, i.e. in the case of
maximum gradient embeddings into trees. We start by describing the
diamond graphs $\{G_k\}_{k=1}^\infty$, and a special labelling of
them that we will use throughout the ensuing arguments. The first
diamond graph $G_1$ is a cycle of length $4$, and $G_{k+1}$ is
obtained from $G_k$ by replacing each edge by a quadrilateral.
Thus $G_k$ has $4^k$ edges and $\frac{2\cdot 4^k+4}{3}$ vertices.
As we have done before, the required lower bound on maximum
gradient embeddings of $G_k$ into trees will be proved if we show
that for every tree $T$ and every non-contractive embedding
$f:G_k\to T$ we have
\begin{eqnarray}\label{eq:goal diamond}
\frac{1}{4^k}\sum_{e\in E(G_k)}\sum_{x\in e} |\nabla f(x)|_\infty
=\Omega\left(k^2\right).
\end{eqnarray}

Note that the inequality~\eqref{eq:goal diamond} is different from
the inequalities that we proved in the case of the cycle and the
path in that the weighting on the vertices of $G_k$ that it
induces is not uniform--- high degree vertices get more weight in
the average in the left-hand side of~\eqref{eq:goal diamond}.

We will prove~\eqref{eq:goal diamond} by induction on $k$. In order
to facilitate such an induction, we will first strengthen the
inductive hypothesis. To this end we need to introduce a useful
labelling of $G_k$. For $1\le i\le k$ the graph $G_k$ contains
$4^{k-i}$ canonical copies of $G_i$, which we index by elements of
$\{1,2,3,4\}^{k-i}$, and denote
$\left\{G^{(k)}_{[\alpha]}\right\}_{\alpha\in \{1,2,3,4\}^{k-i}}$.
These graphs are defined as follows---see Figures 1 and 2 for a
schematic description.
\begin{center}
\centerline{\AffixLabels{\includegraphics[width=1.5in]{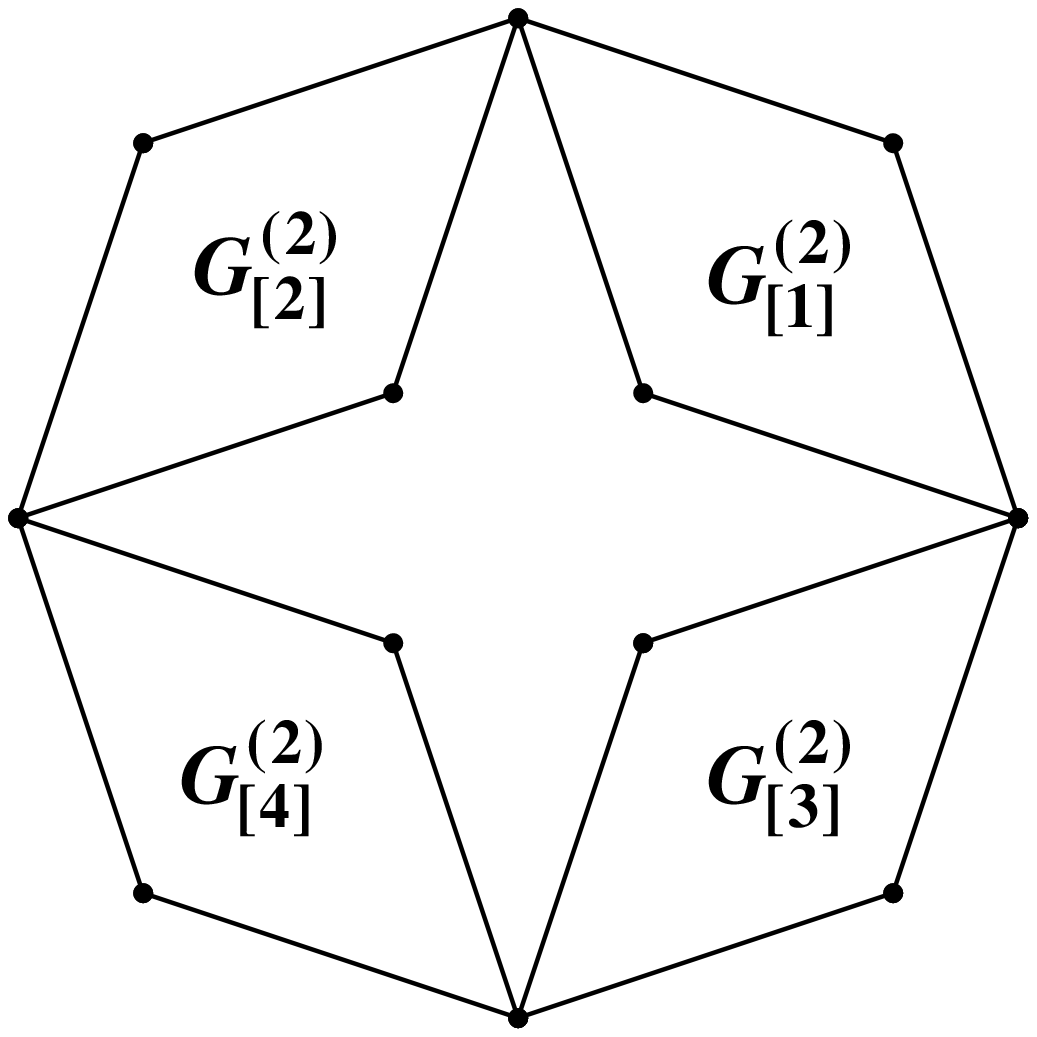}}}

Figure 1: The graph $G_2$ and the labelling of the canonical
copies of $G_1$ contained in it.
\end{center}

%\begin{center}
%\centerline{\AffixLabels{\includegraphics[width=1.5in]{G1.eps}}}
%\end{center}
%{\small Figure 1 : The first diamond graph $G_1$ is a cycle of
%length $4$, with vertices labeled as $\{N,S,E,W\}$. The higher order
%diamond graphs $G_k$ will have certain vertices labeled by strings
%from $\{N,S,E,W\}^k$.}
%\bigskip

\begin{center}
\centerline{\AffixLabels{\includegraphics[height=3.2in,width=3in]{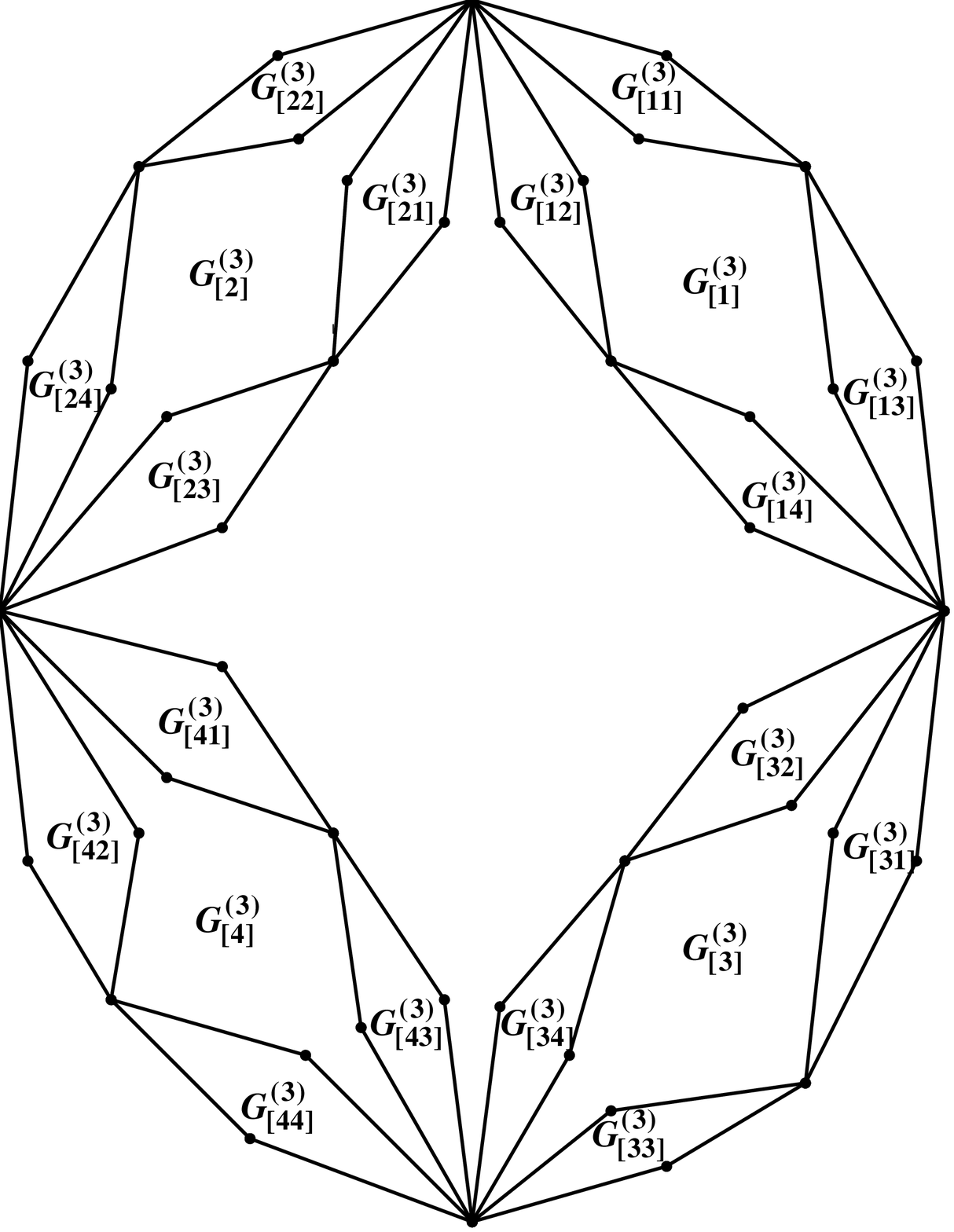}}}

Figure 2: The graph $G_3$ and the induced labelling of canonical
copies of $G_1$ and $G_2$.
\end{center}

Formally, we set $G^{(k)}_{[\emptyset]}=G_k$, and assume inductively
that the canonical subgraphs of $G_{k-1}$ have been defined. Let
$H_1,H_2,H_3,H_4$ be the top-right, top-left, bottom-right and
bottom-left copies of $G_{k-1}$ in $G_k$, respectively. For
$\alpha\in \{1,2,3,4\}^{k-1-i}$ and $j\in \{1,2,3,4\}$ we denote the
copy of $G_i$ in $H_j$ corresponding to $G^{(k-1)}_{[\alpha]}$ by
$G^{(k)}_{[j\alpha]}$.

%Analogously we label the edges of $G_k$ as
%$\{e_\alpha\}_{\alpha\in \{1,2,3,4\}^k}$.

For every $1\le i\le k$ and $\alpha\in \{1,2,3,4\}^{k-i}$ let
$T^{(k)}_{[\alpha]},B^{(k)}_{[\alpha]},L^{(k)}_{[\alpha]},R^{(k)}_{[\alpha]}$
 be the topmost, bottom-most, left-most, and right-most vertices of
$G^{(k)}_{[\alpha]}$, respectively. We will construct inductively
a set of simple cycles $\C_{[\alpha]}$ in $G_{[\alpha]}^{(k)}$ and
for each $C\in \C_{[\alpha]}$ an edge $\e_C\in
E\left(\C_{[\alpha]}\right)$, with the following properties.

\begin{enumerate}
\item The cycles in $\C_{[\alpha]}$ are edge-disjoint, and they all
pass through the vertices $T^{(k)}_{[\alpha]},
B^{(k)}_{[\alpha]},L^{(k)}_{[\alpha]},R^{(k)}_{[\alpha]}$. There
are $2^{i-1}$ cycles in $\C_{[\alpha]}$, and each of them contains
$2^{i+1}$ edges. Thus in particular the cycles in $\C_{[\alpha]}$
form a disjoint cover of the edges in $G^{(k)}_{[\alpha]}$.

\item If $C\in \C_{[\alpha]}$ and $\e_C=\{x,y\}$ then
$d_T(f(x),f(y))\ge \frac{2^{i+1}}{3}-1$.

\item Denote $E_{[\alpha]}=\{\e_C:\ C\in \C_{[\alpha]}\}$ and
$\Delta_i=\bigcup_{\alpha\in \{1,2,3,4\}^{k-i}} E_{[\alpha]}$. The
edges in $\Delta_i$ will be called the {\em designated edges} of
level $i$. For $\alpha\in \{1,2,3,4\}^{k-i}$, $C\in \C_{[\alpha]}$
and $j<i$ let $\Delta_j(C)= \Delta_j\cap E(C)$ be the designated
edges of level $j$ on $C$. Then we require that each of the two
paths $T^{(k)}_{[\alpha]}-L^{(k)}_{[\alpha]}-B^{(k)}_{[\alpha]}$
and $T^{(k)}_{[\alpha]}-R^{(k)}_{[\alpha]}-B^{(k)}_{[\alpha]}$ in
$C$ contains exactly $2^{i-j-1}$ edges from $\Delta_j(C)$.
\end{enumerate}

The construction is done by induction on $i$. For $i=1$ and
$\alpha\in \{1,2,3,4\}^{k-1}$ we let $\C_{[\alpha]}$ contain only
the $4$-cycle $G_{[\alpha]}^{(k)}$ itself. Moreover by
Lemma~\ref{lem:rr} there is and edge $\e_{G_{[\alpha]}^{(k)}}\in
E\left(G_{[\alpha]}^{(k)}\right)$ such that if
$\e_{G_{[\alpha]}^{(k)}}=\{x,y\}$ then $d_T(f(x),f(y))\ge
\frac13$. This completes the construction for $i=1$. Assuming we
have completed the construction for $i-1$ we construct the cycles
at level $i$ as follows. Fix arbitrary cycles $C_1\in
\C_{[1\alpha]}$, $C_2\in \C_{[2\alpha]}$, $C_3\in \C_{[3\alpha]}$,
$C_4\in \C_{[4\alpha]}$. We will use these four cycles to
construct two cycles in $\C_{[\alpha]}$. The first one consists of
the $T^{(k)}_{[\alpha]}-R^{(k)}_{[\alpha]}$ path in $C_1$ which
contains the edge $\e_{C_1}$, the
$R^{(k)}_{[\alpha]}-B^{(k)}_{[\alpha]}$ path in $C_3$ which does
not contain the edge $\e_{C_3}$, the
$B^{(k)}_{[\alpha]}-L^{(k)}_{[\alpha]}$ path in $C_4$ which
contains the edge $\e_{C_4}$, and the
$L^{(k)}_{[\alpha]}-T^{(k)}_{[\alpha]}$ path in $C_2$ which does
not contain the edge $\e_{C_2}$. The remaining edges in $
E(C_1)\cup E(C_2)\cup E(C_3)\cup E(C_4)$ constitute the second
cycle that we extract from $C_1,C_2,C_3,C_4$. Continuing in this
manner by choosing cycles from $\C_{[1\alpha]}\setminus \{C_1\}$,
$\C_{[2\alpha]}\setminus\{C_2\}$, $
\C_{[3\alpha]}\setminus\{C_3\}$, $\C_{[4\alpha]}\setminus\{C_4\}$
and repeating this procedure, and then continuing until we exhaust
the cycles in $\C_{[1\alpha]}\cup \C_{[2\alpha]}\cup
\C_{[3\alpha]}\cup \C_{[4\alpha]}$, we obtain the set of cycles
$\C_\alpha$. For every $C\in \C_\alpha$ we then apply
Lemma~\ref{lem:rr} to obtain an edge $\e_C$ with the required
property.

For each edge $e\in E(G_k)$ let $\alpha\in \{1,2,3,4\}^{k-i}$ be
the unique multi-index such that $e\in
E\left(G_{[\alpha]}^{(k)}\right)$. We denote by $C_i(e)$ the
unique cycle
 in $\C_{[\alpha]}$ containing $e$. We will also denote $\widehat
 e_i(e)=\e_{C_i(e)}$. Finally we let $a_i(e)\in e$ and $b_i(e)\in \widehat
 e_i(e)$ be vertices such that
 $$
d_T(f(a_i(e)),f(b_i(e)))=\max_{\substack{a\in e\\b\in \widehat
 e_i(e)}} d_T(f(a),f(b)).
 $$
Note that by the definition of $\widehat
 e_i(e)$ and the triangle inequality we are assured that
\begin{eqnarray}\label{eq:lower bound for special edge}
d_T(f(a_i(e)),f(b_i(e)))\ge
\frac12\left(\frac{2^{i+1}}{3}-1\right)\ge \frac{2^i}{12}.
 \end{eqnarray}

 Recall that we plan to prove~\eqref{eq:goal
 diamond} by induction on $k$. Having done all of the above preparation, we are now in position to strengthen
\eqref{eq:goal diamond} so as to make the inductive argument
easier. Given two edges $e,h\in G_k$ we write $e\frown_i h$ if
both $e,h$ are on the same canonical copy of $G_i$ in $G_k$,
$C_i(e)=C_i(h)=C$, and furthermore $e$ and $h$ on the same side of
$C$. In other words, $e\frown_i h$ if there is $\alpha\in
\{1,2,3,4\}^{k-i}$ and $C\in \C_{[\alpha]}$ such that if we
partition the edges of $C$ into two disjoint
$T^{(k)}_{[\alpha]}-B^{(k)}_{[\alpha]}$ paths, then $e$ and $h$
are on the same path.

Let $m\in \mathbb N$ be a universal constant that will be
specified later. For every integer $\ell\le k/m$ and any $\alpha
\in \{1,2,3,4\}^{k-m\ell}$ define
$$
L_\ell(\alpha)=\frac{1}{4^{m\ell}}\sum_{e\in E\left(
G_{[\alpha]}^{(k)}\right)}\max_{\substack{i\in \{1,\ldots,\ell\}\\
e\frown_{im} \widehat
e_{im}(e)}}\frac{d_T(f(a_{im}(e)),f(b_{im}(e)))\wedge 2^{im}
}{d_{G_k}(e,\widehat e_{im}(e))+1}.
$$
We also write $L_\ell=\min_{\alpha \in \{1,2,3,4\}^{k-m\ell}}
L_\ell(\alpha)$. We will prove that $L_\ell\ge L_{\ell-1}+c \ell$,
where $c>0$ is a universal constant. This will imply that for
$\ell=\lfloor k/m\rfloor$ we have $L_\ell=\Omega(k^2)$ (since $m$
is a universal constant). By simple arithmetic~\eqref{eq:goal
diamond} follows.

Observe that for every $\alpha\in \{1,2,3,4\}^{k-m\ell}$ we have
\begin{eqnarray*}
 L_\ell(\alpha)&=&\frac{1}{4^m}
\sum_{\beta\in \{1,2,3,4\}^m} \frac{1}{4^{m(\ell-1)}} \sum_{e\in
E\left( G_{[\beta\alpha]}^{(k)}\right)}\max_{\substack{i\in
\{1,\ldots,\ell\}\\ e\frown_{im} \widehat
e_{im}(e)}}\frac{d_T(f(a_{im}(e)),f(b_{im}(e)))\wedge 2^{im}}{d_{G_k}(e,\widehat e_{im}(e))+1}\\
&=&\frac{1}{4^m}\sum_{\beta\in
\{1,2,3,4\}^m}\frac{1}{4^{m(\ell-1)}} \sum_{e\in E\left(
G_{[\beta\alpha]}^{(k)}\right)}\max_{\substack{i\in
\{1,\ldots,\ell-1\}\\ e\frown_{im} \widehat
e_{im}(e)}}\frac{d_T(f(a_{im}(e)),f(
b_{im}(e)))\wedge 2^{im}}{d_{G_k}(e,\widehat e_{im}(e))+1}\\
&\phantom{\le}&+\frac{1}{4^{m\ell}}\sum_{e\in E\left(
G_{[\alpha]}^{(k)}\right)}\max\left\{0,\frac{d_T(f(a_{\ell
m}(e)),f( b_{\ell m}(e)))\wedge 2^{\ell m}}{d_{G_k}(e,\widehat
e_{\ell m}(e))+1}\cdot {\bf 1}_{\{e\frown_{\ell m} \widehat
e_{\ell
m}(e)\}}\phantom{\max_{\substack{i\in \{1,\ldots,\ell-1\}\\
e\frown_{im} \widehat e_{im}(e)}}\frac{d_T(f(a_{im}(e)),f(
b_{im}(e)))}{d_{G_k}(e,\widehat e_{im}(e))+1},0}\right.\\&&\quad\quad\quad\quad\quad\quad\quad\quad\quad\quad\quad\quad\quad\quad\quad-\left.\max_{\substack{i\in \{1,\ldots,\ell-1\}\\
e\frown_{im} \widehat e_{im}(e)}}\frac{d_T(f(a_{im}(e)),f(
b_{im}(e)))\wedge 2^{im}}{d_{G_k}(e,\widehat e_{im}(e))+1}\right\}\\
&=& \frac{1}{4^m}\sum_{\beta\in \{1,2,3,4\}^m}
L_{\ell-1}(\beta\alpha)\\
&\phantom{\le}&+\frac{1}{4^{m\ell}}\sum_{e\in E\left(
G_{[\alpha]}^{(k)}\right)}\max\left\{0,\frac{d_T(f(a_{\ell
m}(e)),f( b_{\ell m}(e)))\wedge 2^{\ell m}}{d_{G_k}(e,\widehat
e_{\ell m}(e))+1}\cdot {\bf 1}_{\{e\frown_{\ell m} \widehat
e_{\ell
m}(e)\}}\phantom{\max_{\substack{i\in \{1,\ldots,\ell-1\}\\
e\frown_{im} \widehat e_{im}(e)}}\frac{d_T(f(a_{im}(e)),f(
b_{im}(e)))}{d_{G_k}(e,\widehat
e_{im}(e))+1},0}\right.\\&&\quad\quad\quad\quad\quad\quad\quad\quad\quad\quad\quad\quad\quad\quad\quad
-\left.\max_{\substack{i\in \{1,\ldots,\ell-1\}\\
e\frown_{im} \widehat e_{im}(e)}}\frac{d_T(f(a_{im}(e)),f(
b_{im}(e)))\wedge 2^{im}}{d_{G_k}(e,\widehat e_{im}(e))+1}\right\}\\
&\ge&L_{\ell-1} +\frac{1}{4^{m\ell}}\sum_{e\in E\left(
G_{[\alpha]}^{(k)}\right)}\max\left\{0,\frac{d_T(f(a_{\ell
m}(e)),f( b_{\ell m}(e)))\wedge 2^{\ell m}}{d_{G_k}(e,\widehat
e_{\ell m}(e))+1}\cdot {\bf 1}_{\{e\frown_{\ell m} \widehat
e_{\ell
m}(e)\}}\phantom{\max_{\substack{i\in \{1,\ldots,\ell-1\}\\
e\frown_{im} \widehat e_{im}(e)}}\frac{d_T(f(a_{im}(e)),f(
b_{im}(e)))}{d_{G_k}(e,\widehat e_{im}(e))+1},0}\right.\\&&\quad\quad\quad\quad\quad\quad\quad\quad\quad\quad\quad\quad\quad\quad\quad-\left.\max_{\substack{i\in \{1,\ldots,\ell-1\}\\
e\frown_{im} \widehat e_{im}(e)}}\frac{d_T(f(a_{im}(e)),f(
b_{im}(e)))\wedge 2^{im}}{d_{G_k}(e,\widehat
e_{im}(e))+1}\right\}.
\end{eqnarray*}

Thus it is enough to show that
\begin{multline}\label{eq:huge goal}
A\stackrel{\mathrm{def}}{=}\frac{1}{4^{m\ell}}\sum_{e\in E\left(
G_{[\alpha]}^{(k)}\right)}\max\left\{0,\frac{d_T(f(a_{\ell
m}(e)),f( b_{\ell m}(e)))\wedge 2^{\ell m}}{d_{G_k}(e,\widehat
e_{\ell m}(e))+1}\cdot {\bf 1}_{\{e\frown_{\ell m} \widehat
e_{\ell
m}(e)\}}\phantom{\max_{\substack{i\in \{1,\ldots,\ell-1\}\\
e\frown_{im} \widehat e_{im}(e)}}\frac{d_T(f(a_{im}(e)),f(
b_{im}(e)))}{d_{G_k}(e,\widehat e_{im}(e))+1}}\right.\\-\left.\max_{\substack{i\in \{1,\ldots,\ell-1\}\\
e\frown_{im} \widehat e_{im}(e)}}\frac{d_T(f(a_{im}(e)),f(
b_{im}(e)))\wedge 2^{im}}{d_{G_k}(e,\widehat
e_{im}(e))+1}\right\}=\Omega(\ell).
\end{multline}

To prove~\eqref{eq:huge goal} denote for $C\in \C_{[\alpha]}$
\begin{multline*}
S_C=\left\{e\in E(C):\ \e_C\frown_{\ell m} e\ \ \mathrm{and}\phantom{\max_{\substack{i\in \{1,\ldots,\ell-1\}\\
e\frown_{im} \widehat e_{im}(e)}}\frac{d_T(f(a_{im}(e)),f(
b_{im}(e)))}{d_{G_k}(e,\widehat e_{im}(e))+1},0}\right.\\ \left.
\max_{\substack{i\in \{1,\ldots,\ell-1\}\\ e\frown_{im} \widehat
e_{im}(e)}}\frac{d_T(f(a_{im}(e)),f( b_{im}(e)))\wedge 2^{im}
}{d_{G_k}(e,\widehat e_{im}(e))+1} \ge \frac12\cdot
\frac{d_T(f(a_{\ell m}(e)),f(b_{\ell m}(e)))\wedge 2^{\ell m}
}{d_{G_k}(e,\widehat e_{\ell m}(e))+1}\right\}.
\end{multline*}
Then using~\eqref{eq:lower bound for special edge} we see that

\begin{eqnarray}
A&\ge& \frac{1}{2\cdot 4^{m\ell}}\sum_{C\in
\C_{[\alpha]}}\sum_{\substack{e\in E(C)\setminus S_C\\
\e_C\frown_{\ell m} e }} \frac{d_T(f(a_{\ell m}(e)),f(b_{\ell
m}(e)))\wedge 2^{\ell m}}{d_{G_k}(e,\widehat e_{\ell m}(e))+1}\nonumber\\
&\ge& \frac{1}{2\cdot 4^{m\ell}}\sum_{C\in
\C_{[\alpha]}}\sum_{\substack{e\in E(C)\\
\e_C\frown_{\ell m} e }} \frac{d_T(f(a_{\ell m}(e)),f(b_{\ell
m}(e)))\wedge 2^{\ell m}}{d_{G_k}(e,\widehat e_{\ell
m}(e))+1}\nonumber-\frac{1}{2\cdot 4^{m\ell}}\sum_{C\in
\C_{[\alpha]}}\sum_{e\in  S_C } \frac{d_T(f(a_{\ell
m}(e)),f(b_{\ell m}(e)))\wedge 2^{\ell m}}{d_{G_k}(e,\widehat e_{\ell m}(e))+1}\nonumber\\
&\ge& \frac{1}{2\cdot 4^{m\ell}}\sum_{C\in
\C_{[\alpha]}}\sum_{i=1}^{2^{m\ell-1}}\frac{2^{m\ell}}{12i}-\frac{1}{2\cdot
4^{m\ell}}\sum_{C\in \C_{[\alpha]}}\sum_{e\in  S_C }
\frac{d_T(f(a_{\ell m}(e)),f(b_{\ell m}(e)))\wedge 2^{\ell m}
}{d_{G_k}(e,\widehat
e_{\ell m}(e))+1}\nonumber\\
&=& \Omega\left(\frac{1}{4^{m\ell}}\cdot |\C_{[\alpha]}|\cdot
2^{m\ell}\cdot m\ell \right)-\frac{1}{2\cdot 4^{m\ell}}\sum_{C\in
\C_{[\alpha]}}\sum_{e\in  S_C } \frac{d_T(f(a_{\ell
m}(e)),f(b_{\ell m}(e)))\wedge 2^{\ell m}}{d_{G_k}(e,\widehat
e_{\ell m}(e))+1}\nonumber\\&=& \Omega(m\ell)-\frac{1}{2\cdot
4^{m\ell}}\sum_{C\in \C_{[\alpha]}}\sum_{e\in S_C }
\frac{d_T(f(a_{\ell m}(e)),f(b_{\ell m}(e)))\wedge 2^{\ell m}
}{d_{G_k}(e,\widehat e_{\ell m}(e))+1}\label{eq:negative on S_C}.
\end{eqnarray}

To estimate the negative term in~\eqref{eq:negative on S_C} fix
$C\in \C_{[\alpha]}$. For every edge $e\in S_C$ (which implies in
particular that $\widehat e_{\ell m}(e)=\e_C$) we fix an integer
$i<\ell$ such that $e\frown_{im} \widehat e_{im}(e)$ and

\begin{multline*}
\frac{2^{im}}{d_{G_k}(e,\widehat e_{im}(e))+1}\ge
\frac{d_T(f(a_{im}(e)),f( b_{im}(e)))\wedge 2^{im}
}{d_{G_k}(e,\widehat e_{im}(e))+1} \ge \frac12\cdot
\frac{d_T(f(a_{\ell m}(e)),f(b_{\ell m}(e)))\wedge 2^{\ell m}
}{d_{G_k}(e,\widehat e_{\ell m}(e))+1}\\ \ge\frac{1}{12}\cdot
\frac{2^{\ell m}}{d_{G_k}(e,\e_C)+1},
\end{multline*}
or
\begin{eqnarray}\label{eq:insert}
d_{G_k}(e,\widehat e_{im}(e))+1\le
2^{(i-\ell)m+4}\left[d_{G_k}(e,\e_C)+1\right].
\end{eqnarray}
We shall call the edge $\widehat e_{im}(e)$ the designated edge
that inserted $e$ into $S_C$. For a designated edge $\e\in E(C)$
of level $im$ (i.e. $\e\in \Delta_{im}(C)$)  we shall denote by
$\mathscr E_C(\e)$ the set of edges of $C$ which $\e$ inserted to
$S_C$. Denoting $D_\e= d_{G_k}(\e,\e_C)+1$ we see
that~\eqref{eq:insert} implies that for $e\in \mathscr E_C(\e)$ we
have
\begin{eqnarray}\label{eq:absolute value}
\big| D_\e- \left[d_{G_k}(e,\e_C)+1\right]\big|\le
2^{(i-\ell)m+4}\left[d_{G_k}(e,\e_C)+1\right].
\end{eqnarray}
Assuming that $m\ge 5$ we are assured that $2^{(i-\ell)m+4}\le
\frac12$. Thus~\eqref{eq:absolute value} implies that
$$
\frac{D_\e}{1+2^{(i-\ell)m+4}}\le d_{G_k}(e,\e_C)+1\le
\frac{D_\e}{1-2^{(i-\ell)m+4}}.
$$
Hence
\begin{eqnarray*}
\sum_{e\in S_C } \frac{d_T(f(a_{\ell m}(e)),f(b_{\ell
m}(e)))\wedge 2^{\ell m} }{d_{G_k}(e,\widehat e_{\ell
m}(e))+1}&\le& \sum_{i=1}^{\ell-1}\sum_{\e\in
\Delta_{im}(C)}\sum_{e\in \mathscr{E}_C(\e)}\frac{2^{\ell
m}}{d_{G_k}(e,\e_C)+1}\\
&\le& 2\sum_{i=1}^{\ell-1}\sum_{\e\in
\Delta_{im}(C)}\sum_{\substack{j\in \mathbb N\\
\frac{D_\e}{1+2^{(i-\ell)m+4}}\le j\le
\frac{D_\e}{1-2^{(i-\ell)m+4}}}}\frac{2^{\ell m}}{j}\\
&=& O(1)\cdot 2^{\ell m}\sum_{i=1}^{\ell-1} |\Delta_{im}(C)|\cdot
\log\left(\frac{1+2^{(i-\ell)m+4}}{1-2^{(i-\ell)m+4}}\right)\\
&=& O(1)\cdot 2^{\ell m}\ell \cdot 2^{(\ell-i) m}\cdot
2^{(i-\ell)m}=O(1)\cdot 2^{\ell m}\ell.
\end{eqnarray*}
Thus, using~\eqref{eq:negative on S_C} we see that
\begin{eqnarray*}
A=\Omega(m\ell)-O(1) \cdot \frac{1}{4^{\ell m}}\cdot
\left|\C_{[\alpha]}\right|2^{m\ell}\ell=\Omega(m\ell)-O(1)\ell=\Omega(\ell),
\end{eqnarray*}
provided that $m$ is a large enough absolute constant.

This completes the proof of the lower bound in
Theorem~\ref{thm:ourmax}.\qed

\section{Monotone clustering problems}

In this section we give some examples which illustrate how certain
monotone clustering problems can be solved efficiently on
ultrametrics. Our arguments are quite flexible, and apply in more
general situations. Before passing to these algorithms, we make a
few general remarks on the framework for monotone clustering that
was discussed in the introduction.

%\subsection{Some remarks on the monotone clustering framework}

 In the definition of monotone clustering
we required that $\Gamma(x,d,P)$ is homogeneous in $d$. One might
wonder whether it is possible to consider also higher orders of
homogeneity, i.e. clustering cost functions $\Gamma$ which satisfy
$\Gamma(x,\lambda d,P)=\lambda^p\Gamma(x,d,P)$ for some $p>1$ (this
occurs, for example, in the $k$-means clustering problem, where the
goal is to find $k$ ``centers" that minimize the sum over the data
points of the squared distance to the closest center). For the proof
of Theorem~\ref{thm:general reduction} to work in this setting we
need a distribution over non-contractive embeddings into
ultrametrics $f:X\to U$ with a polylogarithmic upper bound on the
expected value of $|\nabla f(x)|_\infty^p$. Unfortunately, this is
impossible to achieve in general. Indeed, let
 $f:C_n\to T$ be a random non-contractive embedding of the $n$-cycle into
 trees. Lemma~\ref{lem:rr} implies that there exists an
edge $(x,x+1)\in E(C_n)$ for which $d_T(f(x),f(x+1))\geq
\frac{n}{3}-1$. Thus
$$
\sum_{\{x,y\}\in E(C_n)} d_T(f(x),f(y))^p\ge \frac{n^p}{12^p}.
$$
Taking expectation we see that
$$
\max_{x\in V(C_n)}\E\left[|\nabla f(x)|_\infty^p\right]\ge
\frac{1}{n}\sum_{x\in V(C_n)}\E\left[|\nabla f(x)|_\infty^p\right]\ge
\frac{n^{p-1}}{12^p}.
$$

 We note, however, that the proof of Theorem~\ref{thm:general
 reduction} used the homogeneity of $\Gamma$ in a weak way. In order to get a polylogarithmic reduction to ultrametrics is enough to assume,
 for example,
 that for every $\lambda\ge 1$ we have $\Gamma(x,\lambda
d,P)=O\left(\mathrm{polylog}(n)\right)\cdot\lambda\cdot
\Gamma(x,d,P)$.

Our second remark concerns the fact that the solution space for
monotone clustering problem that was presented in the introduction
was $2^{X\times 2^X}$. This is a huge space, and as we have seen in
Section~\ref{sec:framework}, by setting the clustering cost function
to be $\infty$ on certain possible clustering solutions it is
possible to reduce the size of this space. Additionally, in the
arguments is Section~\ref{sec:framework} the cost function $\Gamma$
ignored the structure of the solution space. Thus in a more generic
formulation of the monotone clustering framework we can assume that
the solution space is some abstract finite set $\mathcal S(X)$. For
example, in our version of the fault-tolerant $k$-median problem we
can take the solution space to be $\binom{X}{k}$.

\subsection{Monotone clustering on ultrametrics via dynamic
programming}\label{sec:dynamic}

We now pass to the design of some monotone clustering algorithms
on ultrametrics. It is a standard fact (see for
example~\cite{BLMN05}) that any ultrametric $(U,d_U)$ can be
represented as follows. There is a graph theoretical tree
$T=(V,E)$ such that $U$ is the set of leaves of $T$. The vertices
of $T$ are labelled by $\Delta: V\to [0,\infty)$ and for every
$u,v\in U$ we have $d_U(u,v)=\Delta(\lca(u,v))$, where $\lca(u,v)$
is the least common ancestor of $u$ and $v$ in $T$. We may, and
will, assume in what follows that every vertex of $T$ is either a
leaf or has exactly two children.

We begin by showing that the fault-tolerant version of the
$k$-median problem described in~\eqref{eq:def fault median} can be
solved exactly on ultrametrics.

\begin{lemma} The minimization of the objective function
in~\eqref{eq:def fault median} can be solved exactly on any
$n$-point ultrametric in time $O(kn^2)$.
\end{lemma}

\begin{proof} Let $(U,d_U)$ be an $n$-point ultrametric and let $T=(V,E)$ be a binary
tree with vertex labels $\Delta:V\to [0,\infty)$ which represents
$U$. We also assume that we are given fault-tolerant parameters
$\{j(u)\}_{u\in U}$. For every $v\in V$ let $T_v$ denote the
subtree of $T$ rooted at $v$. Define for $v\in V$ and $s\in
\{0,\ldots,k\}$
\begin{eqnarray}\label{def:cost*}
\cost^*(v,s)=\min\left\{\sum_{\substack{x\in T_v\cap U\\
j(x)\le s}}d_U\left(x,x_{j(x)}^*(x;d_U)\right):\
x_1,x_2,\ldots,x_s\in T_v\cap U\right\}.
\end{eqnarray}

Our goal is to compute $\cost^*(r,k)$, where $r$ is the root of
$T$. This will be done using dynamic programming. For any leaf
$u\in U$ and $s\in \{0,\ldots,k\}$ define $\cost(u,s)=0$. Let
$v\in V$ be an internal vertex with two children $u,w\in V$.
Define recursively
\begin{multline}\label{eq:dynamic fault}
\cost(v,s)=\min_{t\in
\{0,\ldots,s\}}\Big[\cost(u,t)+\cost(w,s-t)\\+\Delta(v)\cdot \big(
|\{x\in T_u\cap U:\ t<j(x)\le s \}|+|\{x\in T_w\cap U:\
s-t<j(x)\le s \}|\big)\Big].
\end{multline}

A bottom-up computation of the dynamic program
in~\eqref{eq:dynamic fault} computes $\cost(v,s)$ na\"ively in
$O(kn^2)$ time. We will be done if we show that
$\cost(v,s)=\cost^*(v,s)$ for any $v\in V$ and $s\in
\{0,\ldots,k\}$. The fact that $\cost^*(v,s)\le \cost(v,s)$ is
obvious since~\eqref{eq:dynamic fault} computes a feasible
solution of~\eqref{def:cost*} (this fact is proved by a
straightforward induction).

We prove the reverse inequality by induction on $|T_v|$. Let $x_1,\ldots,x_s\in T_v\cap U$ be
such that
$$
\cost^*(v,s)=\sum_{\substack{x\in T_v\cap U\\
j(x)\le s}}d_U\left(x,x_{j(x)}^*(x;d_U)\right).
$$
Let $u,w$ be the children of $v$ in $T$. We may reorder the points
so that for some $t\in \{0,\ldots,s\}$ we have
$\{x_1,\ldots,x_t\}=T_u\cap \{x_1,\ldots,x_s\}$ and
$\{x_{t+1},\ldots,x_s\}=T_w\cap \{x_1,\ldots,x_s\}$. Then

\begin{eqnarray}
&&\!\!\!\!\!\!\!\!\!\!\!\!\!\!\!\!\!\!\cost^*(v,s)=\sum_{\substack{x\in T_v\cap U\\
j(x)\le s}}d_U\left(x,x_{j(x)}^*(x;d_U)\right)\nonumber\\
&=& \sum_{\substack{x\in T_u\cap U\\
j(x)\le t}}d_U\left(x,x_{j(x)}^*(x;d_U)\right)+\sum_{\substack{x\in T_w\cap U\\
j(x)\le s-t}}d_U\left(x,x_{j(x)}^*(x;d_U)\right)\nonumber\\
&\phantom{\le}& + \Delta(v)\cdot \big( |\{x\in T_u\cap U:\
t<j(x)\le s \}|+|\{x\in T_w\cap U:\ s-t<j(x)\le s \}|\big)\label{eq:use ultrametric}\\
&\ge& \cost^*(u,t)
+\cost^*(w,s-t)\nonumber\\&\phantom{\le}&+\Delta(v)\cdot \big(
|\{x\in T_u\cap U:\ t<j(x)\le s \}|+|\{x\in T_w\cap U:\
s-t<j(x)\le s \}|\big)\label{eq:use def cost*}\\
&\ge& \cost(u,t)
+\cost(w,s-t)\nonumber\\&\phantom{\le}&+\Delta(v)\cdot \big(
|\{x\in T_u\cap U:\ t<j(x)\le s \}|+|\{x\in T_w\cap U:\
s-t<j(x)\le s \}|\big)\label{eq:use induction dynamic}\\
&\ge&\cost(v,s),\label{eq:use dynamic cost}
\end{eqnarray}
where in~\eqref{eq:use ultrametric} we used the fact that the tree
$T$ represents the ultrametric $(U,d_U)$, in~\eqref{eq:use def
cost*} we used the definition of $\cost^*(u,t)$ and
$\cost^*(w,s-t)$ given by~\eqref{def:cost*}, in~\eqref{eq:use
induction dynamic} we used the inductive hypothesis, and
in~\eqref{eq:use dynamic cost} we used~\eqref{eq:dynamic fault}.
\end{proof}

Our final result is the proof of Lemma~\ref{lem:PTAS}, which yields
a FPTAS for the $\Sigma\ell_p$ clustering problem on ultrametrics.
We start with the following inequality.

\begin{lemma}\label{lem:norm} Fix $p\ge 1$ and
assume that $a_1\ge a_2\ge\cdots \ge a_n\ge 0$ and
$b_1,\ldots,b_n\ge 0$. Then
$$
\sum_{j=1}^n (a_j^p+b_j^p)^{1/p}\ge \sum_{j=2}^n
a_j+\left(a_1^p+\sum_{j=1}^nb_j^p\right)^{1/p}.
$$
\end{lemma}

\begin{proof} The proof is by induction on $n$, and the inductive
hypothesis simplifies to
\begin{eqnarray}\label{eq:step by step}
\left(a_1^p+\sum_{j=1}^nb_j^p\right)^{1/p}-a_{n+1}\ge
\left(a_1^p+\sum_{j=1}^{n+1}b_j^p\right)^{1/p}-(a_{n+1}^p+b_{n+1}^p)^{1/p}.
\end{eqnarray}
Denote for $x\ge 0$
$$
f(x)=\left(a_1^p+\sum_{j=1}^{n}b_j^p+x\right)^{1/p}-(a_{n+1}^p+x)^{1/p}.
$$
Inequality~\eqref{eq:step by step} is $f(b_{n+1}^p)\le f(0)$, so
it is enough to prove that $f$ is decreasing. But
\begin{eqnarray*}
f'(x)=\frac{1}{p\left(a_1^p+\sum_{j=1}^{n}b_j^p+x\right)^{1-1/p}}-\frac{1}{p\left(a_{n+1}^p+x\right)^{1-1/p}}\le
\frac{1}{p\left(a_1^p+x\right)^{1-1/p}}-\frac{1}{p\left(a_{n+1}^p+x\right)^{1-1/p}}\le
0,
\end{eqnarray*}
since $a_1\ge a_{n+1}$.
\end{proof}

\begin{proof}[Proof of Lemma~\ref{lem:PTAS}]
Let $(U,d_U)$ be an $n$-point ultrametric and let $T=(V,E)$ be a
binary tree with vertex labels $\Delta:V\to [0,\infty)$ which
represents $U$. For $v\in V$, $\ell\in \{0,\ldots,k\}$, $s\in
\{0,\ldots,n\}$ and $t \in [0,\infty)$ define $B^*(v,\ell,s,t)$ to
be the minimum cost according to~\eqref{eq:lp cluster} to cluster
$T_v\cap U$ using $\ell$ sets and centers, when we are allowed to
exclude  $s$ points from $T_v\cap U$, and the most costly cluster
has cost $t$.

We next define a ``pseudo cost" $B(v,\ell,s,t)$ inductively as
follows. If $v$ is a leaf then define
$B(v,1,0,0)=B(v,1,1,0)=B(v,0,1,0)=0$, and for all other values of
$\ell,s,t$ we set $B(v,\ell,s,t)=\infty$. When $v$ has children
$u$ and $w$ define:
\begin{multline*}
B(v,\ell,s,t)=\min\left\{B(u,\ell_1,s_1,t_1)+B(w,\ell_2,s_2,t_2)\phantom{\substack{s_1,r_1,s_2,r_2\in
\{0,\ldots,s\},\\t_1,t_2\in [0,t],\\\ell_1\in
\{0,\ldots,\ell\},\\r_1\le s_1,\\r_2\le s_2,
\\s=s_1+s_2-r_1-r_2,\\\ell=\ell_1+\ell_2,\\t=\max\left\{\left(t_1^p+r_2\Delta(v)^p\right)^{1/p},\
\left(t_2^p+r_1\Delta(v)^p\right)^{1/p}\right\}}}\right.\\\phantom{\le}+\left(t_1^p+r_2\Delta(v)^p\right)^{1/p}-t_1+
\left(t_2^p+r_1\Delta(v)^p\right)^{1/p}-t_2:\left.\substack{s_1,r_1,s_2,r_2\in
\{0,\ldots,s\},\\t_1,t_2\in [0,t],\\\ell_1\in
\{0,\ldots,\ell\},\\r_1\le s_1,\\r_2\le s_2,
\\s=s_1+s_2-r_1-r_2,\\\ell=\ell_1+\ell_2,\\t=\max\left\{\left(t_1^p+r_2\Delta(v)^p\right)^{1/p},\ \left(t_2^p+r_1\Delta(v)^p\right)^{1/p}\right\}}\right\}.
\end{multline*}

With these definition we will prove the following claim by
induction.

\begin{claim}\label{claim:BB*}
For every $v\in T$, $\ell\in \{0,\ldots,k\}$, $s\in
\{0,\ldots,n\}$ and $t \in [0,\infty)$ we have
$$
B^*(v,\ell,s,t)=B(v,\ell,s,t).
$$
\end{claim}

Assuming the validity of Claim~\ref{claim:BB*} for the moment, we
conclude as follows. The dynamic programming algorithm described
above does not suffice since the parameter $t$ takes values in the
range $[0,\infty)$, while we need it to take only $\poly(n)$
values. We fix this issue using an argument which is based on
ideas from~\cite{BCR01}.

Normalize the distances in $U$ so that the minimum distance is
$1$, and denote $\Phi=\diam(U)$. We can clearly assume that $t\le
n\Phi$. Assume first of all that we can ensure that $t\le A=
O\left(\poly(n)\right)$. Once this is achieved then all we need to
do is to apply a standard discretization procedure as follows. Fix
an integer $M>0$  which will be determined presently and let
$A'=\{0,A/M,2A/M,\ldots,A\}$. For $t\in [0,A]$ denote by $\rd(t)$
the rounding of $t$ to its closest value in $A'$. We can now
define a discretized dynamic programming procedure
$B'(v,\ell,s,\tau)$, where $v,\ell,s$ take the same values as in
the definition of $B(v,\ell,s,t)$ and $\tau\in A'$. This is done
by defining as before for a leaf $v\in U$
$B(v,1,0,0)=B(v,1,1,0)=B(v,0,1,0)=0$, and for all other values of
$\ell,s,\tau$ setting $B(v,\ell,s,\tau)=\infty$. When $v$ has
children $u$ and $w$ define:

\begin{multline*}
B'(v,\ell,s,\tau)=\min\left\{\rd\left(\left(\tau_1^p+r_2\Delta(v)^p\right)^{1/p}-\tau_1+
\left(\tau_2^p+r_1\Delta(v)^p\right)^{1/p}-\tau_2\right)\phantom{\substack{s_1,r_1,s_2,r_2\in
\{0,\ldots,s\},\\\tau_1,\tau_2\in [0,t],\\\ell_1\in
\{0,\ldots,\ell\},\\r_1\le s_1,\\r_2\le s_2,
\\s=s_1+s_2-r_1-r_2,\\\ell=\ell_1+\ell_2,\\t=\max\left\{\left(t_1^p+r_1\Delta(v)^p\right)^{1/p},\
\left(\tau_2^p+r_2\Delta(v)^p\right)^{1/p}\right\}}}\right.\\\phantom{\le}+B'(u,\ell_1,s_1,\tau_1)+B'(w,\ell_2,s_2,\tau_2):\left.\substack{s_1,r_1,s_2,r_2\in
\{0,\ldots,s\},\\\tau_1,\tau_2\in A',\\\ell_1\in
\{0,\ldots,\ell\},\\r_1\le s_1,\\r_2\le s_2,
\\s=s_1+s_2-r_1-r_2,\\\ell=\ell_1+\ell_2,\\\tau=
\rd\left(\max\left\{\left(\tau_1^p+r_2\Delta(v)^p\right)^{1/p},\
\left(\tau_2^p+r_1\Delta(v)^p\right)^{1/p}\right\}\right)}\right\}.
\end{multline*}
It is straightforward to check by induction that for any $v\in V$,
$\ell\in \{0,\ldots,k\}$, $s\in \{0,\ldots,n\}$ and $t\in [0,A]$
we have
$$
|B(v,\ell,s,t)-B'(v,\ell,s,\rd(t))|\le \frac{4|T_v|}{M}.
$$
Since the optimal value of the $\Sigma\ell_p$ clustering problem
is at least $1$ (excluding trivial cases), as this is the smallest
distance in $U$, $B'$ will yield an approximation algorithm for
this problem whose multiplicative error is bounded by $1+O(n/M)$.
Taking $M=n/\e$ for some $\e\in (0,1)$ we obtain the required
PTAS.

We therefore need to argue that we can ensure that
$t=O(\poly(n))$. Recall that we can assume that $t\le n\Phi$. Let
$P=\{(x_1,C_1),\ldots,(x_k,C_k)\}$ be the (yet unknown) optimal
solution of the $\Sigma\ell_p$ clustering problem with $k$-centers
on $U$. Let $h$ be the maximum length appearing in the solution,
i.e. $h=\max_{1\le i\le k}\max_{x\in C_i} d_U(x_i,x)$. Fix $\e\in
(0,1)$ and define two ``levels" of the tree $T$ by
$$
L=\left\{v\in V:\ \Delta(v)\le
h<\Delta(\mathrm{parent}(v))\right\},
$$
and
$$
Q=\left\{v\in V:\ \Delta(v)\le \frac{\e h}{n^2}
<\Delta(\mathrm{parent}(v))\right\}.
$$
Let $T'$ be the subtree obtained from $T$ by deleting the subtrees
$\left\{T_v\setminus\{v\}\right\}_{v\in Q}$, and let $U'$ denote
the leaves of $T'$. Equivalently, $U'$ is obtained from $U$ by
contracting all distances smaller that $\e h/n^2$. It is
straightforward to check that $\cost_{U'}(P)\le \cost_U(P)\le
(1+\e)\cost_{U'}(P)$.

Note that for every $v\in L$ the aspect ratio (i.e. the ratio of
the diameter and the shortest distance) of $T_v'\cap U'$ is at
most $n^2/\e$. So, by the above reasoning (in the case of an a
priori polynomial bound on $t$) we can approximate in polynomial
time the value of $B^*(v,\ell,s,t)$ up to a factor $1+O(\e)$. It
remains to ``glue" these approximate solutions to a solution of
the $\Sigma\ell_p$ clustering problem on $T$. This is done by a
(simpler) dynamic programming argument as follows. Denote by
$\widehat T$ the subtree of $T'$ whose root is the same as that of
$T'$ and whose leaves are $L$. For $v\in \widehat T$ let
$C^*(v,\ell)$ be the optimal solution of the $\Sigma\ell_p$
clustering problem on $\widehat T_v$ with $\ell$ centers and
assuming that the largest distance appearing in the solution is at
most $h$. We calculate $C^*(v,\ell)$ by dynamic programming: For
$v\in L$ define $C(v,\ell)=\min_{t} B^*(v,\ell,0,t)$, and if $v$
has two children $u,w$ in $\widehat T$ then
$$
C(v,\ell)=\min \left\{C(u,\ell_1)+C(w,\ell_2):\ \ell_1\in
\{0,\ldots,\ell\},\ \ell_1+\ell_2=\ell \right\}.
$$
A straightforward induction shows that $C^*(v,\ell)=C(v,\ell)$.

The only thing that is left to be explained is how to find the
value $h$. This is done by exhaustive search: We try all the
 $\binom{n}{2}$ possible values of $h$, do the above
procedure for each of them, and take the minimum of the values
that we get.

The proof of Lemma~\ref{lem:PTAS} will be complete once we prove
Claim~\ref{claim:BB*}. We first note that $B^*(v,\ell,s,t)\le
B(v,\ell,s,t)$. This is true because $B(\cdot)$ represents a
feasible solution of $B^*(\cdot)$.  The proof of this fact is by
induction. If $u,w\in V$ are the children of $v$ in $T$ then there
exist $s_1,s_2,t_1,t_2,r_1,r_2,\ell_1,\ell_2$ such that
$$
B(v,\ell,s,t)=B(u,\ell_1,s_1,t_1)+B(w,\ell_2,s_2,t_2)+\left(t_1^p+r_2\Delta(v)^p\right)^{1/p}-t_1+
\left(t_2^p+r_1\Delta(v)^p\right)^{1/p}-t_2,
$$
where $s_1,r_1,s_2,r_2\in \{0,\ldots,s\}$, $t_1,t_2\in [0,t]$, $
\ell_1\in \{0,\ldots,\ell\}$, $r_1\le s_1$, $r_2\le s_2$,
$s=s_1+s_2-r_1-r_2$, $\ell=\ell_1+\ell_2$,
 and $t=\max\left\{\left(t_1^p+r_2\Delta(v)^p\right)^{1/p},\
\left(t_2^p+r_1\Delta(v)^p\right)^{1/p}\right\}$. By the inductive
hypothesis $B(u,\ell_1,s_1,t_1)$ and $B(w,\ell_2,s_2,t_2)$
correspond to feasible solutions of $B^*(\cdot)$ on $T_u\cap U$
and $T_w\cap U$, respectively. Hence $B(v,\ell,s,t)$ corresponds
to the following feasible solution: Take the union of the centers
in $T_u\cap U$ and $T_w\cap U$ and retain all the current clusters
in $T_u\cap U$ and $T_w\cap U$ as is. Next add arbitrary $r_1$
unclustered points from $T_u\cap U$ (from the pool of $s_1$
unclustered points that we are assuming exist in $T_u\cap U$) to
the cluster with the most weight in $T_w\cap U$, and similarly add
$r_2$ unclustered points from $T_w\cap U$ to the cluster with the
most weight in $T_u\cap U$. This creates the required feasible
solution.

We next prove by induction that $B^*(v,\ell,s,t)\ge
B(v,\ell,s,t)$. Consider the clustering solution at which
$B^*(v,\ell,s,t)$ is attained. It corresponds to $s$ excluded
leaves $y_1,\ldots,y_s\in T_v\cap U$, $\ell$ ``centers" $x_1,\ldots,
x_\ell\in (T_v\cap U)\setminus \{y_1,\ldots,y_s\}$ and a partition
$\{C_1,\ldots,C_\ell\}$ of $(T_v\cap U)\setminus
\{y_1,\ldots,y_s\}$ such that
$$
B^*(v,\ell,s,t)=\sum_{j=1}^\ell \left(\sum_{x\in C_j}
d(x,x_j)^p\right)^{1/p}.
$$
%Moreover, assuming without loss of generality that
%$$
%\sum_{x\in C_1} d(x,x_1)^p=\max_{j\in \{1,\ldots,\ell\}}\sum_{x\in
%C_j} d(x,x_j)^p
%$$
%the definition of $B^*(v,\ell,s,t)$ guarantees that
%$$
%\left(\sum_{x\in C_1} d(x,x_1)^p\right)^{1/p}=t.
%$$
By reordering the points we may assume that
$x_1,\ldots,x_{\ell_1}\in T_u$ and
$x_{\ell_1+1},\ldots,x_{\ell_1+\ell_2},\in T_w$ (where
$\ell_2= \ell-\ell_1$). Denote
$$
\left|\left(\bigcup_{j=1}^{\ell_1}C_j\right)\cap
T_w\right|=r_2\quad\mathrm{and}\quad
\left|\left(\bigcup_{j=\ell_1+1}^{\ell_1+\ell_2}C_j\right)\cap
T_u\right|=r_1.
$$
Finally, we may assume that
$$
t_1\stackrel{\mathrm{def}}{=}\sum_{x\in C_1\cap T_u}
d(x,x_1)^p=\max_{j\in \{1,\ldots,\ell_1\}}\sum_{x\in C_j\cap T_u}
d(x,x_j)^p,
$$
and
$$
t_2\stackrel{\mathrm{def}}{=}\sum_{x\in C_{\ell_1+1}\cap T_w}
d(x,x_{\ell_1+1})^p=\max_{j\in
\{\ell_1+1,\ldots,\ell_1+\ell_2\}}\sum_{x\in C_j\cap T_w}
d(x,x_j)^p.
$$
Denote
$$
A_w=\left(\bigcup_{j=1}^{\ell_1}C_j\right)\cap T_w\quad
\mathrm{and} \quad
A_u=\left(\bigcup_{j=\ell_1+1}^{\ell_1+\ell_2}C_j\right)\cap T_u.
$$
We also write $s_1=|\{y_1,\ldots,y_s\}\cap T_u|+r_1$ and
$s_2=|\{y_1,\ldots,y_s\}\cap T_w|+r_2$, so that
$s=s_1+s_2-r_1-r_2$.

Note that by definition
\begin{eqnarray}\label{eq:Tu}
\sum_{j=1}^{\ell_1} \left(\sum_{x\in C_j\cap T_u}
d(x,x_j)^p\right)^{1/p}\ge B^*(u,\ell_1,s_1,t_1),
\end{eqnarray}
and
\begin{eqnarray}\label{eq:Tw}
\sum_{j=\ell_1+1}^{\ell_1+\ell_2}\left( \sum_{x\in C_j\cap T_w}
d(x,x_j)^p\right)^{1/p}\ge B^*(w,\ell_2,s_2,t_2).
\end{eqnarray}
Thus
\begin{eqnarray}
&&\!\!\!\!\!\!\!\!\!\!\!\!\!\!\!\!\!\!\!\!B^*(v,\ell,s,t)=
\sum_{j=1}^{\ell_1}\left[ \sum_{x\in C_j\cap T_u}
d(x,x_j)^p+|C_j\cap
A_w|\Delta(v)^p\right]^{1/p}\nonumber+\sum_{j=\ell_1+1}^{\ell_1+\ell_2}\left[
\sum_{x\in C_j\cap T_w} d(x,x_j)^p+ | C_j\cap
A_u| \Delta(v)^p\right]^{1/p}\nonumber\\
&\ge&
B^*(u,\ell_1,s_1,t_1)+B^*(w,\ell_2,s_2,t_2)+\left(t_1^p+r_2\Delta(v)^p\right)^{1/p}-t_1+
\left(t_2^p+r_1\Delta(v)^p\right)^{1/p}-t_2\label{eq:use step}\\
&\ge&
B(u,\ell_1,s_1,t_1)+B(w,\ell_2,s_2,t_2)+\left(t_1^p+r_2\Delta(v)^p\right)^{1/p}-t_1+
\left(t_2^p+r_1\Delta(v)^p\right)^{1/p}-t_2\label{eq:B
induction}\\
&\ge& B(v,\ell,s,t),\label{eq:def B use}
\end{eqnarray}
where in~\eqref{eq:use step} we used Lemma~\ref{lem:norm} together
with~\eqref{eq:Tu} and~\eqref{eq:Tw}, in~\eqref{eq:B induction} we
used the inductive hypothesis, and in~\eqref{eq:def B use} we used
the definition of $B(\cdot)$. This completes the proof of
Lemma~\ref{lem:PTAS}.
\end{proof}

\end{document}

\remove{ Observe that it follows directly from the recursive
definition in~\eqref{eq:cases1} and~\eqref{eq:cases2} that if
$m_K-x>\sqrt[4]{m_J}$ then necessarily $g_{J_L}(x)$ is given by
the second line in~\eqref{eq:cases2}. So, in this case
$$
g_{J_L}(x)=g_K(x)=\frac{|K|-1}{m_K-x}<\frac{m_J-1}{\sqrt[4]{m_J}}.
$$
}